\documentclass[twocolumn,secnumarabic,amssymb,nobibnotes, aps,prd,superscriptaddress]{revtex4-2}

\usepackage{amsfonts}
\usepackage{amssymb}
\usepackage{amsmath}
\usepackage{dcolumn}
\usepackage{physics}
\usepackage{appendix}
\usepackage{amsthm}

\newtheorem*{proposition}{Proposition}
\usepackage{multirow}
\usepackage{lipsum}
\usepackage{tabularx}
\usepackage{tikz}
\usetikzlibrary{quantikz2}
\usepackage{adjustbox}

\newcommand{\controlled}[1]{\text{C}{#1}}
\newcommand{\nott}{\text{NOT}}

\newcommand{\unitary}{\text{U}}

\newcommand{\id}{\text{I}}
\newcommand{\swapp}{\text{SWAP}}

\newcommand{\su}{\text{SU}}

\newcommand{\fib}{\textbf{\scriptsize{1}}}  
\newcommand{\vac}{\textbf{\scriptsize{0}}}  

\newcommand{\xoroverallerror}{6.64\times 10^{-4}}
\newcommand{\xortargeterror}{6.64\times 10^{-4}}
\newcommand{\xorleakageerror}{3.26\times 10^{-6}}
\newcommand{\xorimage}{xor.png}
\newcommand{\xorscaleimage}{scale.png}
\newcommand{\xorexactimage}{xor-exact.png}

\newcommand{\andoverallerror}{6.64\times 10^{-4}}
\newcommand{\andtargeterrordiff}{6.644\times 10^{-4}}
\newcommand{\andtargeterrorsame}{6.637\times 10^{-4}}
\newcommand{\andleakageerror}{3.99\times 10^{-6}}
\newcommand{\andimage}{and.png}

\newcommand{\andexactimage}{and-exact.png}

\newcommand{\toffolioverallerror}{2.07\times 10^{-3}}
\newcommand{\toffolitargeterror}{8.54\times 10^{-4}}
\newcommand{\toffolileakageerror}{1.62\times 10^{-6}}
\newcommand{\toffoliimage}{toffoli.png}

\newcommand{\toffoliexactimage}{toffoli-injection-exact.png}

\newcommand{\toffolidecooverallerror}{1.90\times 10^{-3}}
\newcommand{\toffolidecotargeterror}{1.77\times 10^{-3}}
\newcommand{\toffolidecoleakageerror}{3.96\times 10^{-6}}
\newcommand{\toffolidecoimage}{toffoli-deco.png}

\newcommand{\toffolidecoexactimage}{toffoli-deco-exact.png}

\newlength{\subfigleng}
\usepackage{graphicx}
\usepackage{subcaption}
\usepackage{hyperref}
\hypersetup{
    colorlinks=true,
    linkcolor=blue,     
    citecolor=blue,     
    urlcolor=blue,      
}

\usepackage{enumitem}
\usepackage{cleveref}

\begin{document}

\title{Optimized Topological Quantum Compilation of Three-Qubit Controlled Gates in the Fibonacci Anyon Model: A Controlled-Injection Approach       
       }

\author{Abdellah Tounsi}
\affiliation{Constantine Quantum Technologies, \\Fr\`{e}res Mentouri University Constantine 1, Ain El Bey Road, Constantine, 25017, Algeria}
\affiliation{Laboratoire de Physique Math\'{e}matique et Subatomique, Fr\`{e}res Mentouri University Constantine 1, Ain El Bey Road, Constantine, 25017, Algeria}%
\author{Nacer Eddine Belaloui}
\affiliation{Constantine Quantum Technologies, \\Fr\`{e}res Mentouri University Constantine 1, Ain El Bey Road, Constantine, 25017, Algeria}
\affiliation{Laboratoire de Physique Math\'{e}matique et Subatomique, Fr\`{e}res Mentouri University Constantine 1, Ain El Bey Road, Constantine, 25017, Algeria}%
\author{Mohamed Messaoud Louamri}
\affiliation{Constantine Quantum Technologies, \\Fr\`{e}res Mentouri University Constantine 1, Ain El Bey Road, Constantine, 25017, Algeria}
\affiliation{Theoretical Physics Laboratory, University of
Science and Technology Houari Boumediene, BP 32 Bab Ezzouar,
Algiers, 16111, Algeria}
\author{Achour Benslama}
\affiliation{Constantine Quantum Technologies, \\Fr\`{e}res Mentouri University Constantine 1, Ain El Bey Road, Constantine, 25017, Algeria}
\affiliation{Laboratoire de Physique Math\'{e}matique et Subatomique, Fr\`{e}res Mentouri University Constantine 1, Ain El Bey Road, Constantine, 25017, Algeria}%
\author{Mohamed Taha Rouabah}
\email{m.taha.rouabah@umc.edu.dz}
\affiliation{Constantine Quantum Technologies, \\Fr\`{e}res Mentouri University Constantine 1, Ain El Bey Road, Constantine, 25017, Algeria}
\affiliation{Laboratoire de Physique Math\'{e}matique et Subatomique, Fr\`{e}res Mentouri University Constantine 1, Ain El Bey Road, Constantine, 25017, Algeria}%

\begin{abstract}
A method, termed controlled-injection, is proposed for compiling three-qubit controlled gates within the non-abelian Fibonacci anyon model. Building on single-qubit compilation techniques with three Fibonacci anyons, the approach showcases enhanced accuracy and reduced braid length compared to the conventional decomposition method for the controlled three-qubit gates. This method necessitates only four two-qubit gates for decomposition, a notable reduction from the conventional five. In conjunction, the study introduces a novel class of controlled three-qubit gates and conducts a numerical simulation of the topological $i$Toffoli gate to validate the approach. In addition, we propose an optimization method for single-qubit gate approximation using novel algebraic relations and numerical methods, including distributed computing.
\end{abstract}

\maketitle

\section{Introduction}
In the previous two decades, there has been a significant interest in topological quantum computation (TQC) due to its potential for scalable and fault-tolerant quantum computing. The later is crucial for fully harnessing the capabilities of quantum systems \cite{Dennis2002, Kitaev2003, Freedman2003, Pachos2012, Lahtinen2017, Stanescu2017, Field2018}. The foundation of TQC relies on the existence of anyons, which were proposed by Jon Magne Leinaas and Jan Myrheim in 1976 \cite{stat} and quantum mechanically formulated by Frank Wilczek in 1982 \cite{wilczek1982}. In (2+1) space-time dimensions, particles can exhibit anyonic statistics, in contrast to the (3+1) dimensions where particles are either bosons or fermions. Consequently, the quantum evolution of anyons moving around each other in an effective two-dimensional space is determined by a topological phase that is independent of the system's dynamics and geometry in the same spirit of Aharonov-Bohm phase of an electron moving around a confined magnetic flux \cite{wilczek1982}. Additionally, within the framework of topological quantum field theory (TQFT), anyon models can be described using braided monoidal category theories \cite{Bakalov2000}. In this context, there exists a correspondence between the topological invariants of knots and the quantum observables of the anyonic system \cite{Witten1989}.
In theory, a class of anyons can be described within the Chern-Simons quantum field theory. This theory is celebrated because of its successful description of the fractional quantum Hall effect (FQHE) \cite{Frhlich1997}. Specifically, the $\su(2)$ non-abelian version of this theory produces the $\su(2)_k$ anyon models, which include the most famous Ising and Fibonacci (Yang-Lee) models for $k=2$ and $k=3$, respectively. It is proven that $\su(2)_k$ anyon models are Turing-complete for $k=3$ and $k\geq 5$ \cite{Freedman2002, Freedman2002-228}, while the $\su(2)_2$ model covers only the Clifford group.
On the other hand, anyons can emerge in some types of topological lattice models. In his seminal work, Kitaev proposed the toric code and quantum doubles as quantum error correction codes that possess topological ground states protected by an energy gap \cite{Kitaev2003}. In this framework, abelian and non-abelian anyons emerge as excitations and serve as quantum information processing agents. Recently, a modified version of the surface code has been simulated in a superconducting quantum processor, providing evidence of the non-abelian statistics of Ising anyons \cite{google2023}. At the same time, a quantum double code based on the Dihedral group $D_4$ implemented on a trapped-ion quantum processor demonstrated experimental observation of non-abelian anyon statistics \cite{iqbal2023creation}. Moreover, the experimental results from various condensed matter systems provide compelling evidence in support of the realization of topological quantum computing systems in the near future. Specifically, it has been shown experimentally that the quasi-particles in the fractional quantum Hall systems exhibit anyonic statistics \cite{Willett2013, Nakamura2020}. Furthermore, a recent study suggests the possibility of detecting Majorana zero modes in semiconductor-superconductor heterostructure devices \cite{PhysRevB.107.245423}.

However, the compilation of quantum circuits in the TQC framework is not generally trivial. Although the Fibonacci anyon model ensures universality and is dense in the special unitary group, significant efforts have been devoted to achieving an efficient and optimal compilation scheme. Numerous algorithms are available to map $\su(2)$ quantum gates to topological braiding operations within Fibonacci anyons, including the Solovay-Kitaev algorithm \cite{Kitaev_1997}, the brute-force algorithm \cite{Field2018, Rouabah2021}, the evolutionary algorithm \cite{PhysRevB.87.054414}, and deep reinforcement learning \cite{Moro2021}. Additionally, an asymptotically optimal and systematic algorithm based on ring theory has been specifically designed for certain $\su(2)$ quantum gates with Fibonacci anyons \cite{PhysRevLett.112.140504}. Moreover, a recently developed generic Monte Carlo approach has been introduced to compile near-optimal braid words at the single-qubit level in $\su(2)_k$ anyon models \cite{PRXQuantum.2.010334}.

Additionally, it is necessary to compile two-qubit gates, such as the CNOT gate, to ensure universality. In this context, several procedures have been proposed. Specifically, an injection method was introduced by Bonesteel et. al. \cite{Bonesteel2005} to  compose two-qubit gates from single-qubit gates of Fibonacci anyons. Another iterative procedure has been designed to systematically generate entangling two-qubit Fibonacci anyon braids \cite{PhysRevA.93.052328}. Eventhough the compilation of any single-qubit gate along with CNOT gate is sufficient to demonstrate the quantum computing universality of a given anyon model, the implementation of larger quantum circuits may require reduced compilation schemes to decrease processing time. Specifically, the Toffoli gate, or the controlled-controlled NOT gate, can be decomposed by concatenation of five two-qubit controlled gates \cite{PhysRevA.88.010304}. However, the pertinent query revolves around the feasibility of reducing this requisite.
In this context, a general compilation scheme based on dense encoding was proposed to construct controlled-controlled phase gate for a wide range of anyon models using six successive gates acting on three qubits densely encoded on eight anyons \cite{physreva.84.012332}. This method does not address a wide range of controlled-controlled gates such as the CNOT and Deutsch gates.
In this study, we present a novel procedure that we call controlled-injection to construct a class of controlled three-qubit gates within the Fibonacci anyon model. The controlled-injection method offers the advantage of reducing the complexity of the braids compared to the conventional decomposition method. We demonstrate our finding by comparing the resulting $i$Toffoli gates using both approaches. To ensure an accurate comparison using the most optimal compilation possible, we employ the brute-force algorithm and implement several optimizations based on algebraic and numerical techniques, given the computational complexity of the brute-force algorithm. Furthermore, the controlled-injection method introduces unusual quantum logic gates that can be used in specific applications.\\

This paper begins by introducing the fundamentals of quantum computing with Fibonacci anyons, along with essential notations. We then delve into the key aspects of compiling single-qubit gates within the Fibonacci model, using the brute-force algorithm. Thereafter, we present the controlled injection method, supplemented by a new class of controlled three-qubit gates. We conclude our study with a comparative analysis between the newly introduced method and the conventional decomposition approach.

\section{Fibonacci Model}
\label{sec:fibonacci-model}

The Fibonacci anyon model is the simplest universal anyon model among $\su(2)_k$ family. Fibonacci anyons can theoretically emerge as quasiparticles in the FQHE at $12/5$ filling factor \cite{SLINGERLAND2001229}. They can also appear as defects in the so called string-net (Levin-Wen) lattice models \cite{PhysRevB.71.045110, PhysRevB.86.165113}. In the Fibonacci model, we have only one non-trivial anyonic charge, referred to as $\fib$, along with the trivial vacuum charge, $\vac$. The non-trivial Fibonacci fusion rule is defined as follows \cite{bonderson, Rouabah2021, Simon2023}:
\begin{align}
   \fib \times \fib = \vac + \fib.
\end{align}
The dimension of the fusion space $\mathcal{F}$ in terms of the number of Fibonacci anyons follows the famous Fibonacci series such that $\dim(\mathcal{F}_{n+2})=\dim(\mathcal{F}_{n}) +\dim(\mathcal{F}_{n+1})$, where $n$ is the number of anyons, as shown in Table \ref{tab:fibonacci_dim}. Notice that a qubit can be represented minimally by three or four anyons. While two anyons introduce a trivial braid operation, three anyons form a qubit with overall charge $\fib$ and correspond to the minimum qubit representation with Fibonacci anyons. Four anyons with an overall charge $\text{\vac}$ form also a qubit. Representing a qubit with more than four anyons is not recommended since it provokes leakage, as proven for all anyon models \cite{Ainsworth2011}.

To compute the matrix representation of the braid generators of a given set of anyons, the fusion and rotation matrices must be computed. Solving the pentagon identities reveals one non-trivial fusion matrix $F$ in Fibonacci model given by \cite{bonderson, Pachos2012}:
\begin{align}
    F^{\fib}_{\fib \fib \fib}&=
    \begin{pmatrix}
        (F^{\fib}_{\fib \fib \fib})^\vac_\vac & (F^{\fib}_{\fib \fib \fib})^\vac_\fib \\
        (F^{\fib}_{\fib \fib \fib})^\fib_\vac & (F^{\fib}_{\fib \fib \fib})^\fib_\fib
    \end{pmatrix}
=
    \begin{pmatrix}
        \frac{1}{\phi}        & \frac{1}{\sqrt{\phi}} \\
        \frac{1}{\sqrt{\phi}} & -\frac{1}{\phi}
    \end{pmatrix},
\end{align}
where $\phi$ is the golden ratio. Then, solving the hexagon identities gives the right-handed and the left-handed solutions of the rotation matrix $R$ \cite{bonderson, Pachos2012}:
\begin{align}
    R_{\fib \fib} &=  
    \begin{pmatrix}
        R_{\fib \fib}^{\vac}      & R_{\fib \fib}^{\vac \fib}  \\
        R_{\fib \fib}^{\fib \vac} & R_{\fib \fib}^{\fib}
    \end{pmatrix} =
    \begin{pmatrix}
        e^{\pm i4\pi/5}  & 0                \\
        0                & -e^{\pm i2\pi/5}
    \end{pmatrix}.
\end{align}
Here, the $(+)$ and $(-)$ signs in the exponents refer to right-handed and left-handed twisting respectively. The off-diagonal components of $R_{\fib \fib}$ are null because exchanging two particles preserves the fusion outcome.

Braiding, a fundamental operation in anyon models, consists of exchanging two adjacent anyons. The braiding operation between the $n$-th and $(n+1)$-th particles is denoted by the $\sigma_n$ braid operator.
In this work, all the necessary braid operators matrix representations are calculated using a systematic numerical method developed in \cite{tounsi2023systematic}. Fortunately, the braid operator $\sigma_3$ in the case of four anyons is identical to the operator $\sigma_1$. That implies that all single-qubit gates acting on three anyons yield the same evolution when applied to four anyons.\\
\begingroup
\squeezetable
\begin{table}
	\centering
	\caption{\label{tab:fibonacci_dim} This table shows the size of the fusion space $\mathcal{F}_{n}$ in function of the number of anyons $n$. $\dim(\mathcal{F}_{n}^{\vac})$ is the dimension of the fusion space given that the total charge is $\vac$, while $\dim(\mathcal{F}_{n}^{\fib})$ is the size of the fusion space given that the overall charge is $\fib$. Finally, $\dim(\mathcal{F}_{n})$ is the total size of the fusion space.}
	\begin{ruledtabular}
    \begin{tabular}{cccc}
		$n$ & $\dim(\mathcal{F}_{n}^{\vac})$ & $\dim(\mathcal{F}_{n}^{\fib})$ & $\dim(\mathcal{F}_{n})$\\
		\hline
		1 & 0 & 1 & 1\\
		2 & 1 & 1 & 2\\
		3 & 1 & 2 & 3\\
		4 & 2 & 3 & 5\\
		$n$ & $\sum_{i=1,2}\dim(\mathcal{F}_{n-i}^\vac)$ & $\sum_{i=1,2}\dim(\mathcal{F}_{n-i}^\fib)$ & $\sum_{i=1,2}\dim(\mathcal{F}_{n-i})$
	\end{tabular}
    \end{ruledtabular}
\end{table}
\endgroup

The dimension of the fusion space of anyons differs from that of the Hilbert space of qubits. For instance, simulating two qubits requires a fusion space of size four or larger. In the context of Fibonacci models, two practical options arise. The first is to use five anyons prepared with an overall charge of $\fib$, resulting in a fusion space of five dimensions. The second is to allocate three anyons to each qubit, resulting in five and eight fusion states depending on whether the overall charge is $\vac$ or $\fib$, respectively. The former option, termed dense encoding, is considered optimal in terms of resource utilization. The latter, called sparse encoding, is compatible with the picture of quantum circuits decomposed into separate qubits. A general scheme for gate compilation within the dense encoding framework is explored in \cite{physreva.84.012332}. However, in this work, we opt for sparse encoding, aligning with the mainstream \cite{Bonesteel2005, Hormozi2007, Field2018}. Additionally, this study demonstrates several advantages of this convention.

\section{Compiling Single-Qubit Gates}
\label{sec:compiling-single-qubit-gates}
The algebraic group generated by Fibonacci braiding operations is densely mapped to the $\su(2)$ group and is polynomially equivalent to a quantum circuit \cite{Freedman2002}. In the context of three Fibonacci anyons, the matrix representation of any Fibonacci braid sequence $\hat B$ assumes a general form \cite{Hormozi2007}:
\begin{align}
\label{eq:u(b)-formula}
    B
    &=
    \left(
    \begin{array}{cc|c}
        \multicolumn{2}{c|}{\multirow{2}{*}{$\pm \text{e}^{-iW(\hat B)\pi/10}$ [\su(2)]}} & \\
        \multicolumn{2}{c|}{}                                                       & \\
        \hline
        &  & \text{e}^{i3W(\hat B)\pi/5}
    \end{array}
    \right),
\end{align}
where $W(\hat B)$, the winding number, is defined as the sum of the powers of the braid sequence. Consequently, the global phase of $B$  in both blocks is determined solely by the winding number, up to a $\pm$ sign. While the global phase factor may not be of significant importance when compiling single-qubit gates, it manifests measurable effects in the design of two and three-qubit controlled gates. Furthermore, the phase difference between the two independent sectors of fusion charges $\fib$ and $\vac$ gains relevance when executing mixing operations between two qubits \cite{Hormozi2007}. As a result, it is advisable to employ a global phase-independent distance metric, given that the formula in Eq. \eqref{eq:u(b)-formula} precisely determines the global phase.\\

It has been established that the weave group, which is a subgroup of the braid group, is also dense to $\su(2)$ \cite{Simon2006}. Weaves are defined as sequences of braids wherein a single anyon navigates around other stationary anyons. In diagrammatic terms, there exists a solitary warp strand that weaves around the remaining weft strands. The primary objective of this method is to simplify the implementation of braiding circuits. In scenarios involving three anyons, weaving sequences are restricted to braiding operations with even powers.\\

In scholarly discourse, numerous algorithms have been proposed to identify the most accurate braid sequence of a certain length that approximates a targeted unitary gate. The Solovay-Kitaev algorithm is a seminal method in this context, as it demonstrates that an approximation of a given quantum gate with a braid sequence up to the desired level of accuracy can be achieved efficiently in polylogarithmic time \cite{Hormozi2007}. As per the Solovay-Kitaev theorem, the relationship between the error $\epsilon$ and the braiding length $L$ is as follows:
\begin{align}
    L = \text{poly}( \log(1/\epsilon)).
\end{align}
Here, $\epsilon$ is proportional to any properly defined distance metric between the braid unitary matrix and the target unitary matrix in the space of unitary matrices. In this study, we utilize the spectral distance metric \cite{Bonesteel2005, khatri2019} to quantify the error $\epsilon$:
\begin{align}
    \label{eq:distance-metric}
    \mathcal{D}(U_1, U_2) = \sqrt{\text{maxEigenvalue}(AA^\dagger)},
\end{align}
such that $A$ is the difference between $U_1$ and $U_2$ after eliminating the global phases.\\

Despite the fact that the brute-force algorithm imposes an exponential demand on computational resources, it yields an optimal fundamental approximation. This suffices to provide a proof of concept for the gates introduced in subsequent sections. To render the brute-force algorithm more practical, its performance has been augmented with a number of optimizations.
Firstly, for the optimization of linear calculations, it is advantageous to map the $\su(2)$ components of the elementary braid matrices \eqref{eq:u(b)-formula} to quaternions. This approach is computationally more efficient as it reduces the number of parameters and necessitates fewer real number multiplications. Secondly, the cyclicity of the braid matrix powers should be taken into account. Specifically, $\sigma_1^{10} = \sigma_2^{10} = \id$. As a result, in the context of weaves, it is adequate to span only sequences of the form $\sigma_i^p$, where $i=1,2$ and $p=2,4,6,8$.
Thirdly, the braid group enforces algebraic relations that can effectively reduce the search space. This is achieved by leveraging the similarity relation that associates each braid sequence with its counterpart, wherein $\sigma_1$ and $\sigma_2$ are permuted. Namely,
\begin{align}
    \sigma_1^{p_n} \sigma_2^{q_n} \cdots \sigma_1^{p_1} \sigma_2^{q_1}
    = \Gamma^\dagger \sigma_2^{p_n} \sigma_1^{q_n} \cdots \sigma_2^{p_1} \sigma_1^{q_1} \Gamma,
\end{align}
where $p_i$ and $q_i$ are integers and $\Gamma = \sigma_1 \sigma_2 \sigma_1$. The physical interpretation of the similarity relation, along with its proof, are elaborated in Appendix \ref{app:FBF}. Furthermore, we illustrate in Appendix \ref{app:proposition} that if the target gate is Hermitian, it suffices to span merely half of the search space. In essence, a Hermitian target gate $H$ introduces a symmetry in the metric space since for each braid matrix $B$:
\begin{align}
    \mathcal{D}(B, H) = \mathcal{D}(B^\dagger, H).
\end{align}

From a computational perspective, the utilization of a high-performance programming language can lead to substantial savings in terms of computational time, energy, memory, and development duration \footnote{In this study, the \texttt{C++} language, recognized as one of the contemporary high-performance languages \cite{10.1145/3136014.3136031}, is employed.}.
Furthermore, a linear acceleration is achievable by partitioning the search space across available processing units. Collectively, all algebraic and numerical optimizations significantly reduced the search time.

\section{Compiling topological Conditional Three-Qubit Gates using controlled-injection method}
\label{sec:compiling-three-qubit-gates}
\begin{figure}[t]
    \centering
    \includegraphics[width=\linewidth]{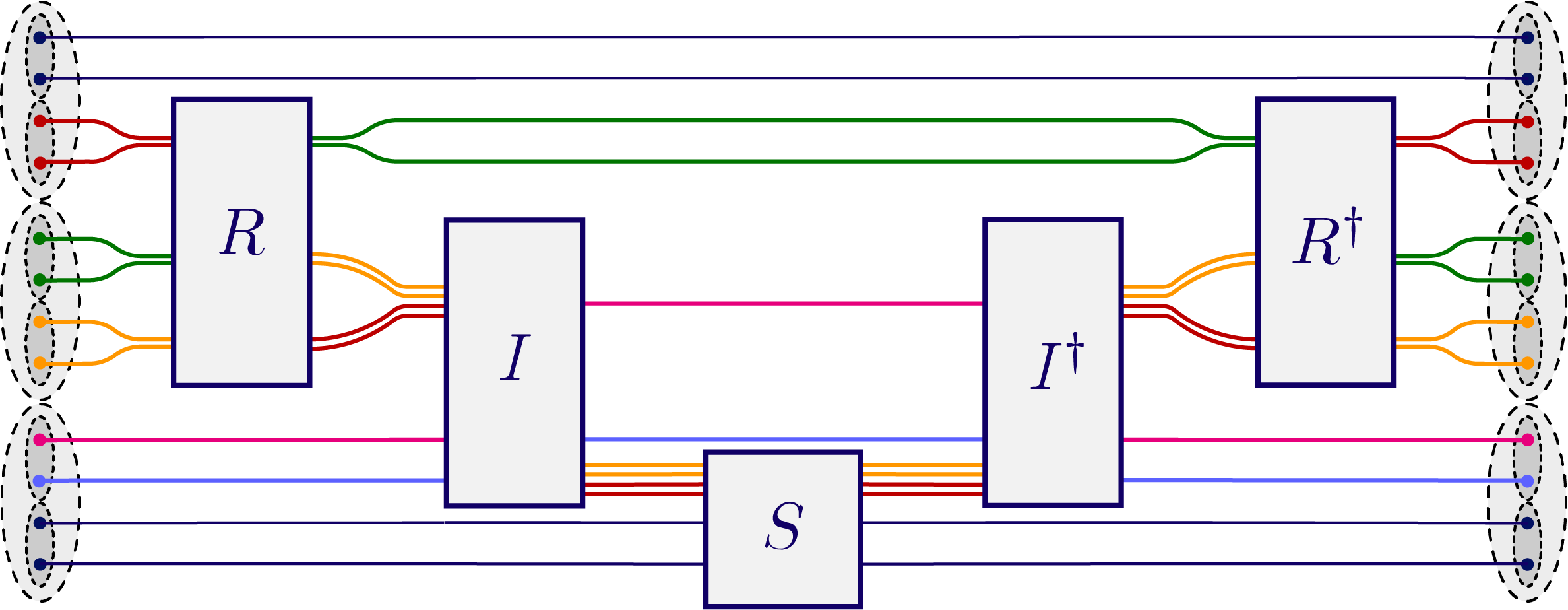}
    \caption{General scheme of the proposed three-qubit gate by the controlled-injection method. The two upper qubits are the controlling qubits, while the lower qubit is the controlled qubit. The $R$ operation prepares the controlling qubits while the $S$ operation is the target operation which is applied whenever the injected four anyons have a non-trivial overall topological charge.}
    \label{fig:three-qubit-gate}
\end{figure}

We introduce a class of quantum three-qubit gates $\mathbb{M}(R, S)$ that perform controlled logic operations inherently compatible with the structure of the Fibonacci model. We call this method as the \textit{controlled-injection} method. As shown in Fig. \ref{fig:three-qubit-gate}, the general form of this class of three-qubit gates encodes qubits in four anyon groups. This convention is necessary for this method and is the simplest encoding whose topological charge $\vac$ sector has two dimensions. The topological sector of charge $\vac$  serves the purpose of the exact implementation of the SWAP gate since the braiding of two $\vac$ charges is trivial. The latter is frequently needed to implement controlled gates between non-adjacent qubits.
In addition, the introduced $\mathbb{M}(R, S)$ three-qubit gate involves three-anyon gates labeled $R, I$ and $S$ which act as $\su(2)\oplus \unitary(1)$ operations, where the $\su(2)$ part acts on the topological charge $\fib$ sector while \unitary(1) part acts on the topological sector of charge $\vac$ as shown in Eq. \eqref{eq:u(b)-formula}.
We compile these three-anyon gates by weaving only one input anyon as explained in Section \ref{sec:compiling-single-qubit-gates}. However, we are allowed to weave pairs and groups of anyons keeping in mind only the fusion outcomes of each group. In this case, the three-anyon gates will affect only the fusion state of the fusion outcomes regardless to the constituent anyons. As shown in Fig. \ref{fig:three-qubit-gate}, the gate $R$ takes pairs of anyons as inputs while weaving only the upper pair, the gate $I$ injects the upper grouped two pairs in yellow and red colors into the lower strand, and the $S$ gate weaves the upper grouped two pairs returning them to the same strand position.
In general, the $\mathbb{M}(R, S)$ gate works in such a way that the $R$ gate prepares the controlling qubits, so we call it the \textit{initialization gate}, while the $S$ gate is the target operation to be applied on the controlled qubit, then, we call it the \textit{target gate}. We call the intermediary identity gate the \textit{injection gate}. To understand precisely the mechanism of this gate, we will introduce two variations of the controlled-injection three-qubit gates: $\mathbb{M}(I, S)$ and $\mathbb{M}(\nott, S)$ whose initialization gates are the identity and the $\nott$ gate respectively.\\

\subsection{The $\mathbb{M}(I,S)$ controlled-injection gate}
Let us consider the initialization gate to be the identity operation i.e $R=I$. Notice that in the case of a four anyons qubit of overall topological charge $\vac$, the two constituent anyon pairs fuse simultaneously to $\vac$ if the state is $\ket{0}$:
\begin{align}
    \ket{0} = \ket{((\fib, \fib)_\vac, (\fib, \fib)_\vac)_\vac},
\end{align}
 and fuse simultaneously to $\fib$ if the state is $\ket{1}$:
\begin{align}
    \ket{1} = \ket{((\fib, \fib)_\fib, (\fib, \fib)_\fib)_\vac}.
\end{align}
Therefore, the identity initialization gate will have no effect but injecting the lower pair of anyons from the upper qubit into the middle qubit. If one and only one qubit of the two controlling qubits is in the state $\ket{0}$, the middle two pairs will have different topological charges with overall topological charge $\fib$. In contrast, when the controlling qubits are both in the state $\ket{0}$ or both in the state $\ket{1}$, the middle two pairs will have the same topological charges, and since the initialization gate approximates the identity in $\mathbb{M}(I,S)$, the overall topological charge of the middle two pairs will remain $\vac$. We summarize the truth table of this gate in Tab. \ref{tab:truth-table-M(I,S)}, where the target gate $S$ is applied if the controlling qubits have opposite logic states. As a result, the exact matrix representation of $\mathbb{M}(I, S)$ should be given by:

\begin{align}
    \left(
    \begin{array}{cc|cc|cc|cc}
    1 & 0 & 0 & 0 & 0 & 0 & 0 & 0\\
    0 & 1 & 0 & 0 & 0 & 0 & 0 & 0\\
    \hline
    0 & 0 & \multicolumn{2}{c|}{\multirow{2}{*}{\huge $S$}} & 0 & 0 & 0 & 0\\
    0 & 0 & & & 0 & 0 & 0 & 0\\
    \hline
    0 & 0 & 0 & 0 & \multicolumn{2}{c|}{\multirow{2}{*}{\huge $S$}} & 0 & 0\\
    0 & 0 & 0 & 0 & & & 0 & 0\\
    \hline
    0 & 0 & 0 & 0 & 0 & 0 & 1 & 0\\
    0 & 0 & 0 & 0 & 0 & 0 & 0 & 1
    \end{array}
    \right)
\end{align}

In conclusion, the $\mathbb{M}(I,S)$ gate can be represented as the application of three consecutive two-qubit controlled gates. Specifically, it can be expressed as follows:
\begin{align}
    \mathbb{M}(I, S) &= \left( \text{CNOT} \otimes I \right) \left(I \otimes \text{CS}\right) \left( \text{CNOT} \otimes I \right).
\end{align}
This representation provides a clear understanding of the operation of the $\mathbb{M}(I,S)$ gate in terms of standard quantum gates.
\begingroup
\begin{table}
    \centering
    \caption{\label{tab:truth-table-M(I,S)} Truth table of $\mathbb{M}(I, S)$ gate.}
    \begin{ruledtabular}
    \begin{tabular}{c c c c c c}
        \multicolumn{3}{c}{Input} & \multicolumn{3}{c}{Output}\\
        \hline
        qubit 1 & qubit 2 & qubit 3 & qubit 1 & qubit 2 & qubit 3 \\
        \hline
        $\ket{0}$ & $\ket{0}$ & $\ket{\eta}$ & $\ket{0}$ & $\ket{0}$ & $\ket{\eta}$ \\
        $\ket{0}$ & $\ket{1}$ & $\ket{\eta}$ & $\ket{0}$ & $\ket{1}$ & $S\ket{\eta}$\\
        $\ket{1}$ & $\ket{0}$ & $\ket{\eta}$ & $\ket{1}$ & $\ket{0}$ & $S\ket{\eta}$\\
        $\ket{1}$ & $\ket{1}$ & $\ket{\eta}$ & $\ket{1}$ & $\ket{1}$ & $\ket{\eta}$
    \end{tabular}
\end{ruledtabular}
\end{table}
\endgroup

\subsubsection{Numerical simulation of $\mathbb{M}(I, \nott)$}

In practice, the initialization, injection, and target gates are approximated to a predetermined braid length. This approximation process inevitably leads to a degree of inaccuracy and leakage. To assess the efficacy of the $\mathbb{M}(I, \nott)$ gate, we conducted a numerical simulation with the target gate chosen as the typical NOT gate, specifically $\pm iX$, the special unitary version of the NOT gate.
The initial step involves identifying the weave approximation of the $\pm \id$ gate for the initialization and injection gates, under the stipulation that the weft strand's initial tip takes the upper rank while its end tip takes the lower rank. Concurrently, we search for the weave approximation of the $S$ gate, which acts on three strands, weaving the upper strand without changing its final rank.
Through the optimized brute-force approach, we derived relevant weaving sequences by setting a fixed braid length of 48 braid operators and accounting for global phases. It is noteworthy that there exists a multitude of weaving sequences that exhibit optimal accuracy. In our numerical implementation, we select the weaving sequences that combine to provide the best accuracy for the $\mathbb{M}(I,\pm iX)$ gate.
The sequences pertinent to this case are delineated in Table \ref{tab:weaving-sequences-M(I,S)}.

\begingroup
\squeezetable
    \begin{table*}
    \centering
    \caption{\label{tab:weaving-sequences-M(I,S)} Braid sequence and accuracy of the necessary three-strand gates to approximate the $\mathbb{M}(I, iX)$ gate with an overall error of $\xoroverallerror$ and leakage $\epsilon_L = \xorleakageerror$.}
    \begin{ruledtabular}
    \begin{tabular}{lllcc}
        $\mathbb{M}(R, S)$ Gates & $\mathbb{M}(I, iX)$ Gates & Weave Sequence & Length & Error\\
        \hline
        & & &\\
        R & $\id$ &
        \scriptsize{$\sigma_1^{-1}\sigma_1^{-2}\sigma_2^{2}\sigma_1^{4}\sigma_2^{-2}\sigma_1^{-4}\sigma_2^{-2}\sigma_1^{2}\sigma_2^{2}\sigma_1^{4}\sigma_2^{4}\sigma_1^{2}\sigma_2^{-4}\sigma_1^{-2}\sigma_2^{2}\sigma_1^{-2}\sigma_2^{-2}\sigma_1^{2}\sigma_2^{-2}\sigma_2^{-1}$} 
        & 48 & $1.51\times 10^{-3}$\\
        I & $\id$ &
        \scriptsize{$\sigma_1^{1}\sigma_1^{2}\sigma_2^{2}\sigma_1^{4}\sigma_2^{4}\sigma_1^{4}\sigma_2^{2}\sigma_1^{4}\sigma_2^{2}\sigma_1^{2}\sigma_2^{-2}\sigma_1^{2}\sigma_2^{-2}\sigma_1^{2}\sigma_2^{2}\sigma_1^{2}\sigma_2^{-2}\sigma_1^{2}\sigma_2^{-2}\sigma_1^{2}\sigma_2^{-1}$} & 48 & $1.51\times 10^{-3}$\\
        S & $iX$ &
        \scriptsize{$\sigma_2^{1}\sigma_2^{4}\sigma_1^{-2}\sigma_2^{-2}\sigma_1^{-2}\sigma_2^{-4}\sigma_1^{-2}\sigma_2^{2}\sigma_1^{2}\sigma_2^{-4}\sigma_1^{2}\sigma_2^{4}\sigma_1^{-2}\sigma_2^{4}\sigma_1^{-2}\sigma_2^{-4}\sigma_1^{-2}\sigma_2^{-2}\sigma_2^{-1}$} 
        & 48 & $8.55\times 10^{-4}$\\
        &&&
    \end{tabular}
    \end{ruledtabular}
    \end{table*}
\endgroup

The subsequent phase involves the conversion of these three-strand gates into a six-strand braiding circuit, as depicted in Fig. \ref{fig:three-qubit-gate}.
Ultimately, employing the relevant matrix representations of the five elementary braid operations, a systematic numerical method developed in \cite{tounsi2023systematic} is utilized to obtain the representation of the approximated $\mathbb{M}(I,iX)$ gate, as illustrated in Fig. \ref{fig:M(I,iX)-simulation}.
The accuracy is quantified by calculating the error using the distance metric defined in Eq. \eqref{eq:distance-metric}. The overall error of this $\mathbb{M}(I,iX)$ approximation is computed as $\xoroverallerror$. The error measured solely on the controlled gate (the $S\equiv \pm iX$ block in the $\mathbb{M}(I,iX)$ matrix) is approximately $\xortargeterror$. This value is commensurate with the error computed on the $\pm iX$ approximation itself. The overall error is comparable to the error on the target because the $\mathbb{M}(R,S)$ is always accurate when the controlled gate is not applied since trivial braids are involved, and the target gate contributes significantly to the inaccuracy.\\

The leakage error, denoted as $\epsilon_L$, can also be computed. This error represents the quantity of information that the simulated gate inevitably transmits to non-computational states. It is calculated using a modified form of the spectral distance. Specifically, the leakage induced by a given gate $U$ of $d$ dimensions is defined as:
\begin{align}
	\label{eq:leakage}
	\epsilon_L = 1-\sqrt{\text{minEigenvalue} (UU^\dagger)}.
\end{align}
In this equation, the second term yields the minimum factor by which the matrix $U$ can alter the norm of a quantum state \cite{Field2018}. The leakage amount measured on the simulated $\mathbb{M}(I,iX)$ gate is approximately $\xorleakageerror$. It is important to note that the initialization and injection gates are anticipated to be primarily responsible for this leakage of information.\\

\begin{figure}[t]
     \centering
     \begin{subfigure}[b]{\subfigleng}
         \centering
         \includegraphics[width=\textwidth]{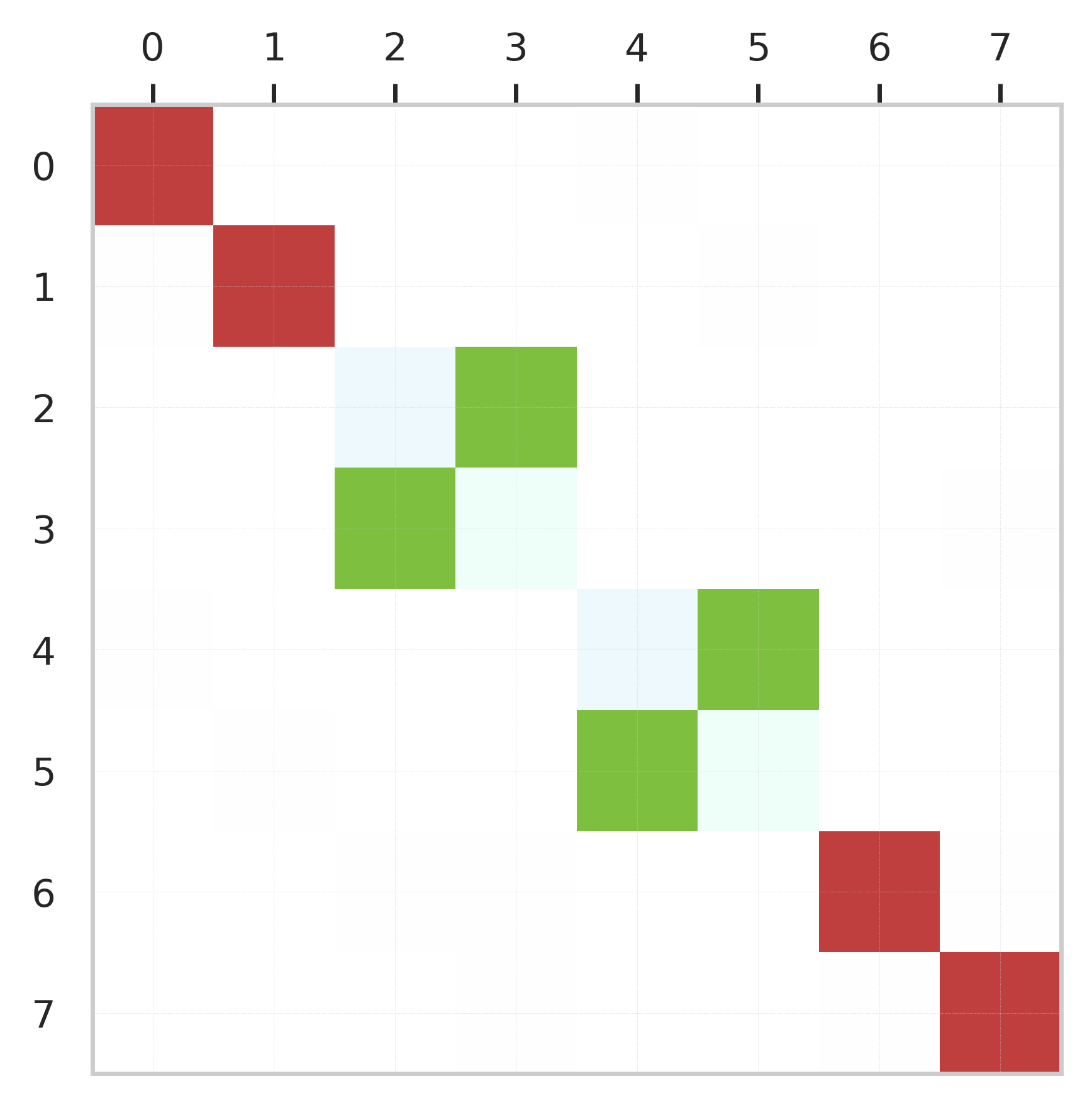}
         \caption{\label{fig:M(I,iX)-sim}}
     \end{subfigure}
     \hfill
     \begin{subfigure}[b]{\subfigleng}
         \centering
         \includegraphics[width=\textwidth]{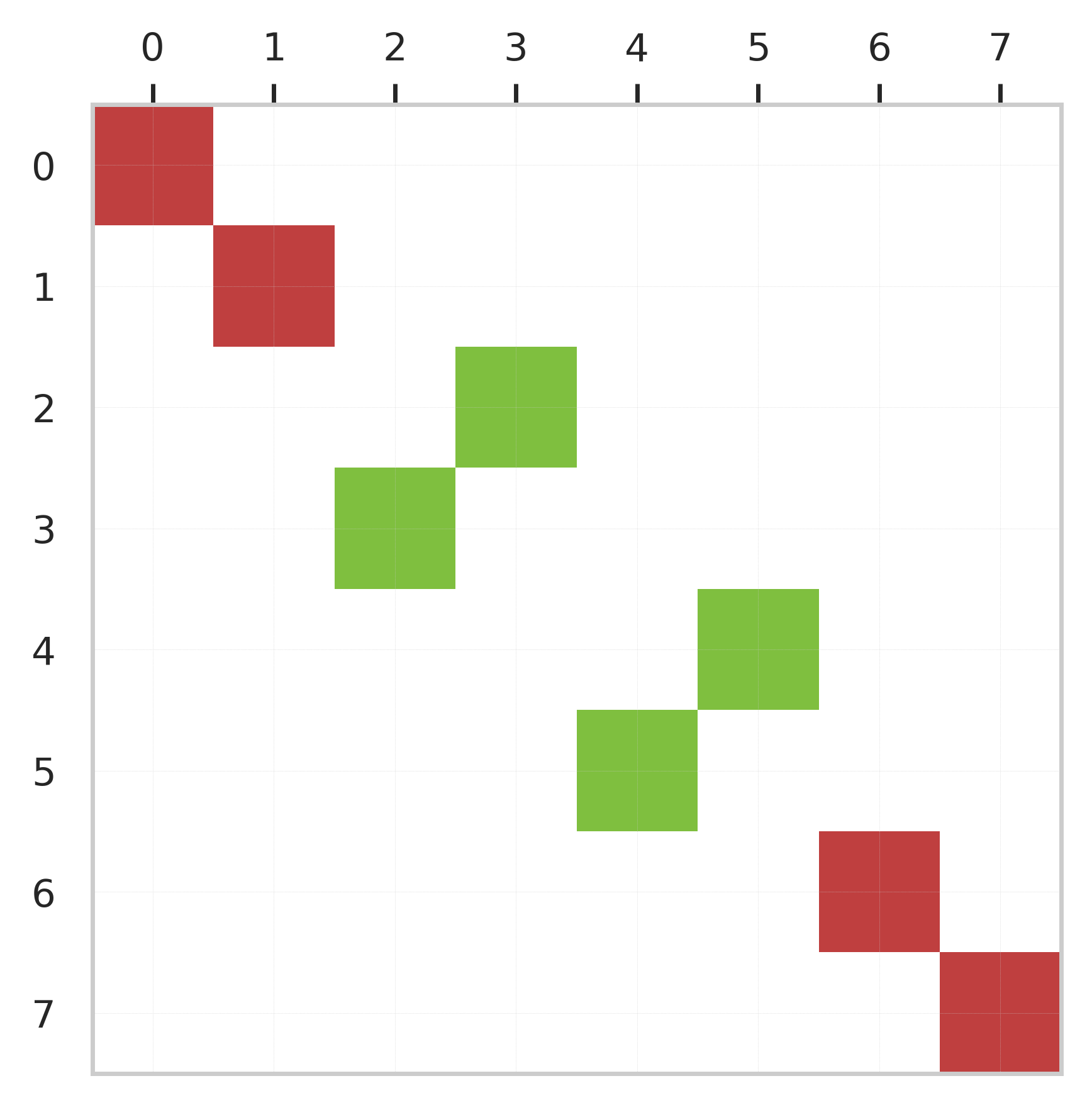}
         \caption{\label{fig:M(I,iX)-exact}}
     \end{subfigure}
     \hfill
     \begin{subfigure}[b]{\subfigleng}
        \centering
        \includegraphics[width=\textwidth]{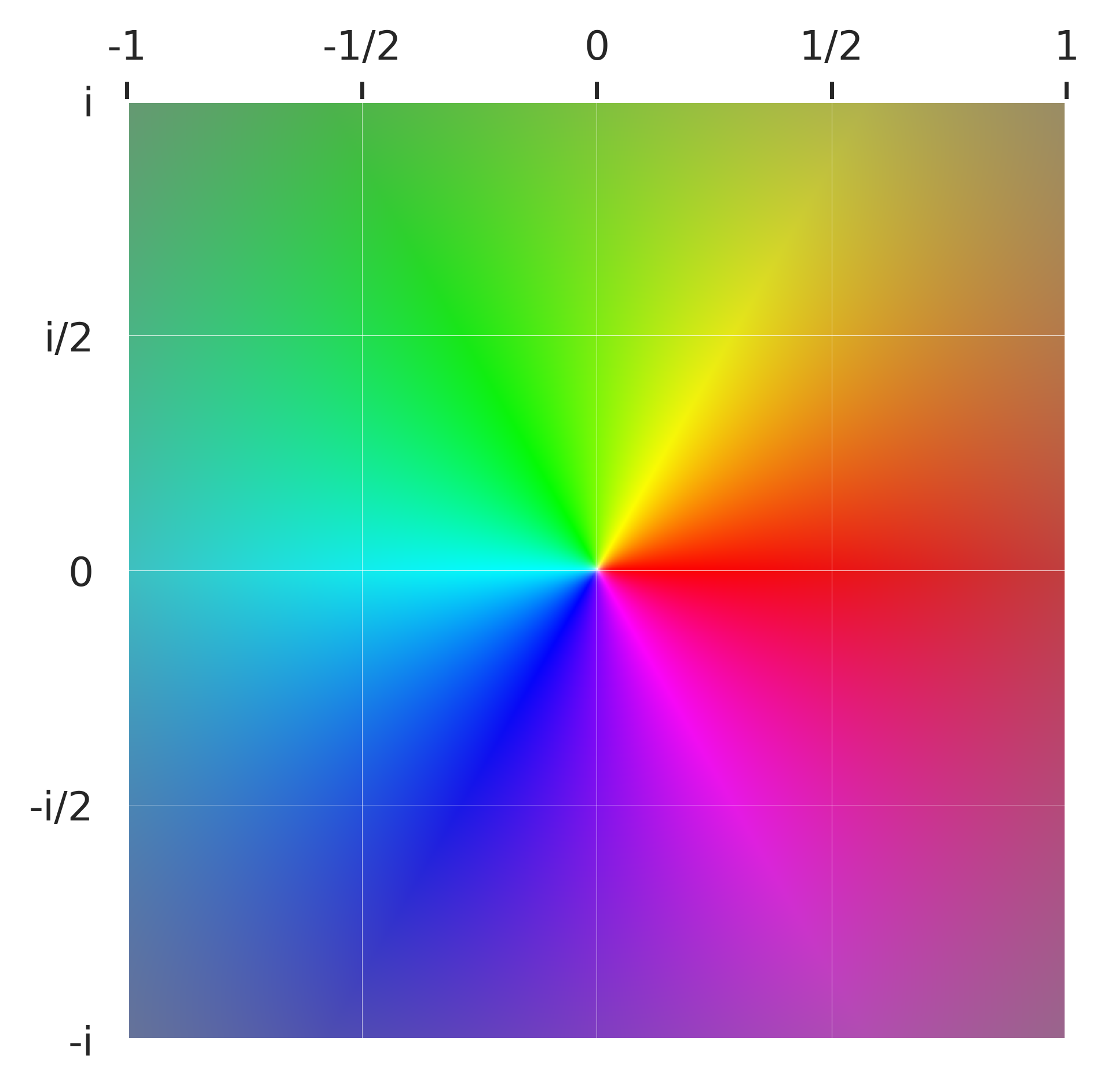}
        \caption{\label{fig:scale}}
    \end{subfigure}
    \caption{(a) Matrix representation of the approximated $\mathbb{M}(I,iX)$ in the computational basis. The full braid matrix is $89\times 89$ in the sector of charge $\vac$ while the sector $\fib$ is irrelevent because the overall charge cannot be $\fib$. (b) Exact $\mathbb{M}(I,iX)$ matrix representation. (c) The RGB color map is designed to map each complex number, with an amplitude less than one, to a specific color. This mapping process involves a linear transformation of the radial component into spectral colors. Additionally, the amplitude is mapped into luminosity in the vicinity of zero and into a saturation profile near the identity, following a Gaussian distribution.}
    \label{fig:M(I,iX)-simulation}
\end{figure}

\subsection{The $\mathbb{M}(\nott,S)$ controlled-injection gate}

Let us now examine the scenario where the initialization gate assumes the role of the NOT gate, i.e., $R=\pm iX$. The complex phase $\pm i$ is essential for maintaining gates in the $\su(2)$ group and preventing complex phases in the sector $\fib$, as illustrated in Eq. \ref{eq:u(b)-formula}. To comprehend the effect of the initialization gate, one must revisit how the NOT gate operates on the fusion states of three Fibonacci anyons \cite{tounsi2023systematic}.
When both controlling gates are in the state $\ket{1}$, the initialization gate reverses the overall topological charge of the middle two pairs from the charge $\vac$ to the charge $\fib$:
\begin{align}
\nott \ket{((\fib, \fib)_\vac, \fib)_\fib} = \ket{((\fib, \fib)_\fib, \fib)_\fib},
\end{align}
since the fusion states $\ket{((\fib, \fib)_\vac, \fib)_\fib}$ and $\ket{((\fib, \fib)_\fib, \fib)_\fib}$ are the only logic states in case of three fibonacci anyons in the sector of charge $\fib$.
In other cases, the action of the initialization gate is analogous to the $R=I$ case and is trivial.
Therefore, the target gate will only be operational when neither of the controlling qubits are in the $\ket{0}$ state. The truth table of $\mathbb{M}(iX, S)$ should align with that shown in Tab. \ref{tab:truth-table-M(iX,S)}. Consequently, the relevant matrix representation of the $\mathbb{M}(NOT, S)$ is as follows:
\begin{align}
    \left(
    \begin{array}{cc|cc|cc|cc}
    1 & 0 & 0 & 0 & 0 & 0 & 0 & 0\\
    0 & 1 & 0 & 0 & 0 & 0 & 0 & 0\\
    \hline
    0 & 0 & \multicolumn{2}{c|}{\multirow{2}{*}{\huge $S$}} & 0 & 0 & 0 & 0\\
    0 & 0 & & & 0 & 0 & 0 & 0\\
    \hline
    0 & 0 & 0 & 0 & \multicolumn{2}{c|}{\multirow{2}{*}{\huge $S$}} & 0 & 0\\
    0 & 0 & 0 & 0 & & & 0 & 0\\
    \hline
    0 & 0 & 0 & 0 & 0 & 0 & \multicolumn{2}{c}{\multirow{2}{*}{\huge $S$}}\\
    0 & 0 & 0 & 0 & 0 & 0 & &
    \end{array}
    \right)
\end{align}

\begingroup
    \begin{table}[h]
    \centering
    \caption{\label{tab:truth-table-M(iX,S)} Truth table of the $\mathbb{M}(iX, S)$ Gate.}
    \begin{ruledtabular}
    \begin{tabular}{cccccc}
        \multicolumn{3}{c}{Input} & \multicolumn{3}{c}{Output}\\
        \hline
        qubit 1 & qubit 2 & qubit 3 & qubit 1 & qubit 2 & qubit 3 \\
        \hline
        $\ket{0}$ & $\ket{0}$ & $\ket{\eta}$ & $\ket{0}$ & $\ket{0}$ & $\ket{\eta}$ \\
        $\ket{0}$ & $\ket{1}$ & $\ket{\eta}$ & $\ket{0}$ & $\ket{1}$ & $S\ket{\eta}$\\
        $\ket{1}$ & $\ket{0}$ & $\ket{\eta}$ & $\ket{1}$ & $\ket{0}$ & $S\ket{\eta}$\\
        $\ket{1}$ & $\ket{1}$ & $\ket{\eta}$ & $\ket{1}$ & $\ket{1}$ & $S\ket{\eta}$
    \end{tabular}
    \end{ruledtabular}
    \end{table}
\endgroup

\subsubsection{Numerical simulation of $\mathbb{M}(\nott,\nott)$}

As an illustrative instance, we consider $S=\pm iX$ and set the length of the three-anyon braid gates to 48. The initial step involves identifying the compilation of $\pm iX$ and $\pm I$, ensuring that the weaving process commences from the upper strand and concludes at the lower strand.
Subsequently, we seek an approximate weave for the $S=\pm iX$ gate, which operates on three strands, initiating the weaving process from the upper strand and terminating at the same strand. Utilizing a an optimized brute-force approach, we derive the necessary weave sequences, as presented in Table \ref{tab:M(iX,iX)}.
These sequences are procured by spanning all combinations of the optimal individual braid sequences to enhance the accuracy of the $\mathbb{M}(iX, iX)$ gate.\\

\begingroup
\squeezetable
    \begin{table*}
    \centering
    \caption{Braid approximation of the necessary three-strand gates to compile the $\mathbb{M}(iX, iX)$ gate with an overall error of $\andoverallerror$ and leakage $\epsilon_L = \andleakageerror$.}
    \begin{ruledtabular}
    \begin{tabular}{lllcc}
        $\mathbb{M}(R,S)$ Gates & $\mathbb{M}(iX,iX)$ Gates & Weave sequence & Length & Error\\
        \hline
        &&&\\
        R &$iX$ &
        \scriptsize{$\sigma_1^{-1}\sigma_2^{2}\sigma_1^{-4}\sigma_2^{2}\sigma_1^{-2}\sigma_2^{2}\sigma_1^{-2}\sigma_2^{4}\sigma_1^{-2}\sigma_2^{-2}\sigma_1^{-2}\sigma_2^{4}\sigma_1^{-2}\sigma_2^{-2}\sigma_1^{2}\sigma_2^{2}\sigma_1^{-2}\sigma_2^{-2}\sigma_1^{-2}\sigma_2^{-2}\sigma_1^{-2}\sigma_2^{1}$} 
        & 48 & $8.55\times 10^{-4}$\\
        I & $\id$ &
        \scriptsize{$\sigma_1^{1}\sigma_1^{2}\sigma_2^{2}\sigma_1^{4}\sigma_2^{4}\sigma_1^{4}\sigma_2^{2}\sigma_1^{4}\sigma_2^{2}\sigma_1^{2}\sigma_2^{-2}\sigma_1^{2}\sigma_2^{-2}\sigma_1^{2}\sigma_2^{2}\sigma_1^{2}\sigma_2^{-2}\sigma_1^{2}\sigma_2^{-2}\sigma_1^{2}\sigma_2^{-1}$} 
        & 48 & $1.51\times 10^{-3}$\\
        S&$iX$ &
        \scriptsize{$\sigma_2^{1}\sigma_2^{4}\sigma_1^{-2}\sigma_2^{-2}\sigma_1^{-2}\sigma_2^{-4}\sigma_1^{-2}\sigma_2^{2}\sigma_1^{2}\sigma_2^{-4}\sigma_1^{2}\sigma_2^{4}\sigma_1^{-2}\sigma_2^{4}\sigma_1^{-2}\sigma_2^{-4}\sigma_1^{-2}\sigma_2^{-2}\sigma_2^{-1}$} 
        & 48 & $8.55\times 10^{-4}$\\
        &&&
    \end{tabular}
    \end{ruledtabular}
    \label{tab:M(iX,iX)}
    \end{table*}
\endgroup

The matrix representation of the simulated $\mathbb{M}(iX,iX)$ gate is depicted in Fig. \ref{fig:M(iX,iX)-simulation}. The computed error distance of the braid approximation is found to be $\andoverallerror$. However, the error of the controlled gate in the level of $S$ blocks is approximately $\andtargeterrordiff$ if the controlling qubits are either in the $\ket{01}$ state or in the $\ket{10}$ state, while it is computed to be $\andtargeterrorsame$ if the controlling qubits are in the $\ket{11}$ state. The amount of leakage is around $\andleakageerror$.\\

\begin{figure}[t]
     \centering
     \begin{subfigure}[b]{\subfigleng}
         \centering
         \includegraphics[width=\textwidth]{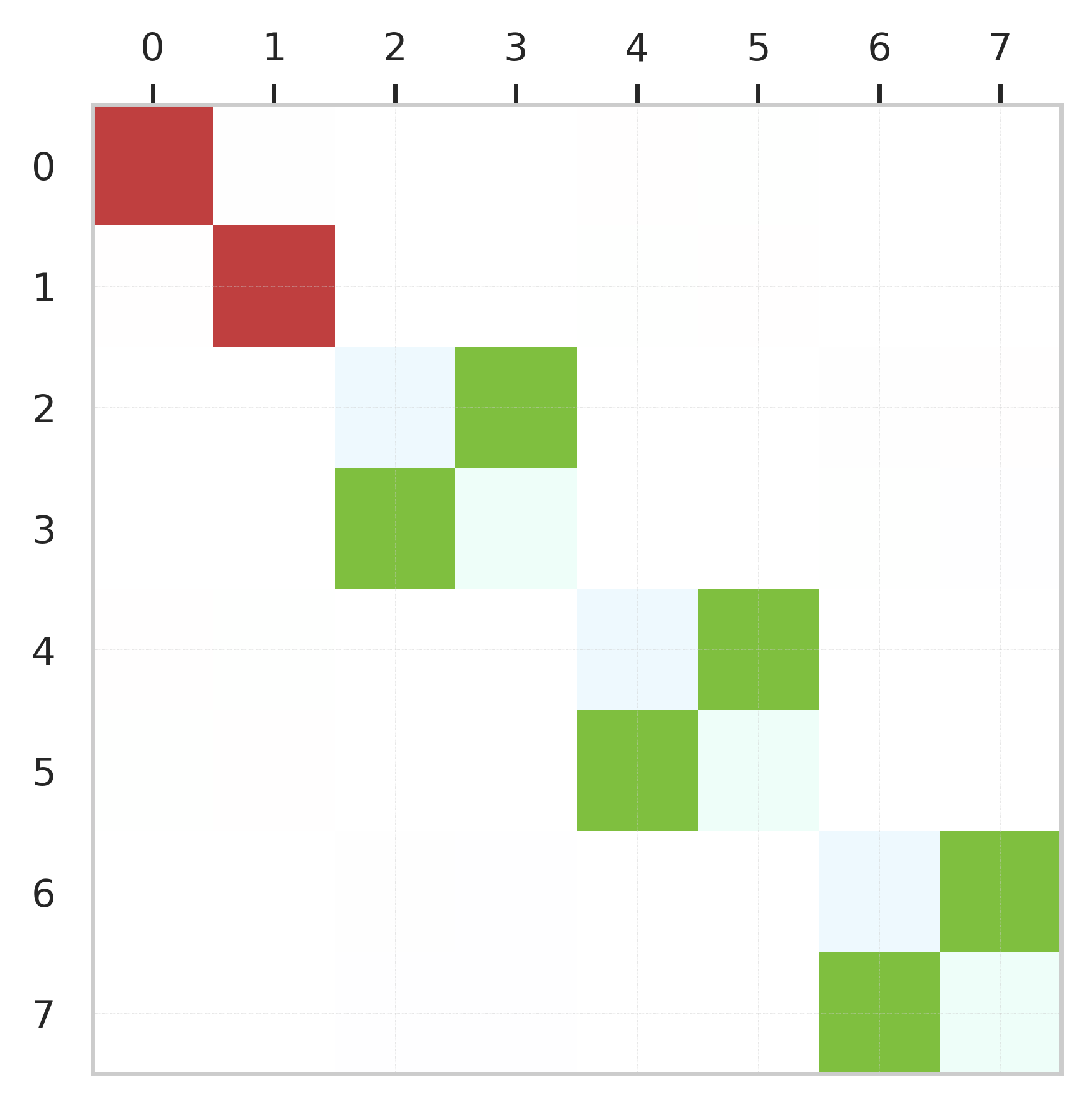}
         \caption{\label{fig:M(iX,iX)-sim}}
     \end{subfigure}
     \begin{subfigure}[b]{\subfigleng}
         \centering
         \includegraphics[width=\textwidth]{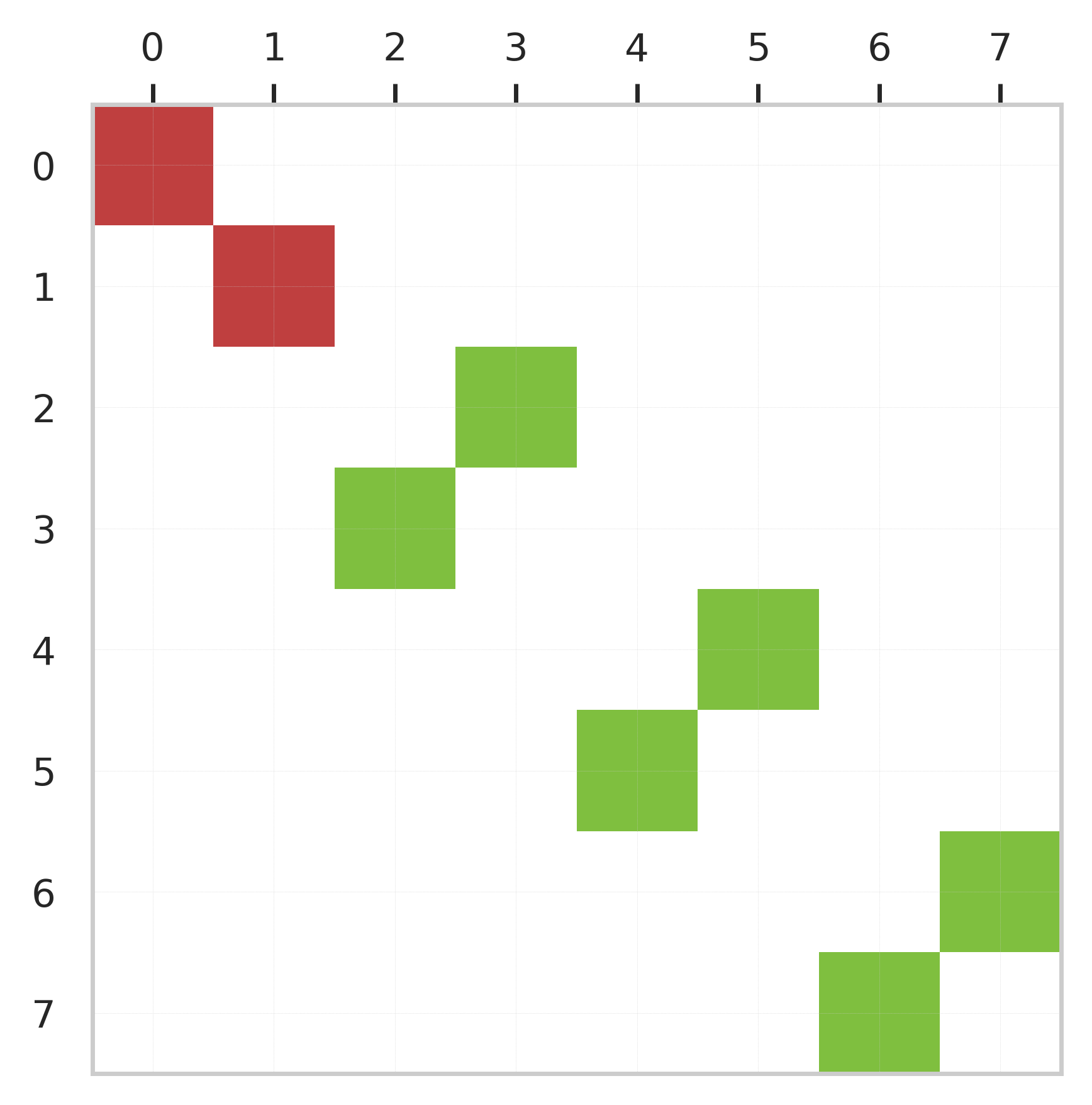}
         \caption{\label{fig:M(iX,iX)-exact}}
     \end{subfigure}
        \caption{\label{fig:M(iX,iX)-simulation} (a) Matrix representation of the
        approximated $\mathbb{M}(iX,iX)$ as expressed in the basis of the computational states. (b) Exact $\mathbb{M}(iX,iX)$ matrix representation. The color map is depicted in Fig. \ref{fig:scale}.}
\end{figure}

\subsection{Controlled-Controlled-S gate with controlled-injection method}

\begin{figure}[t]
    \centering
    \includegraphics[width=\linewidth]{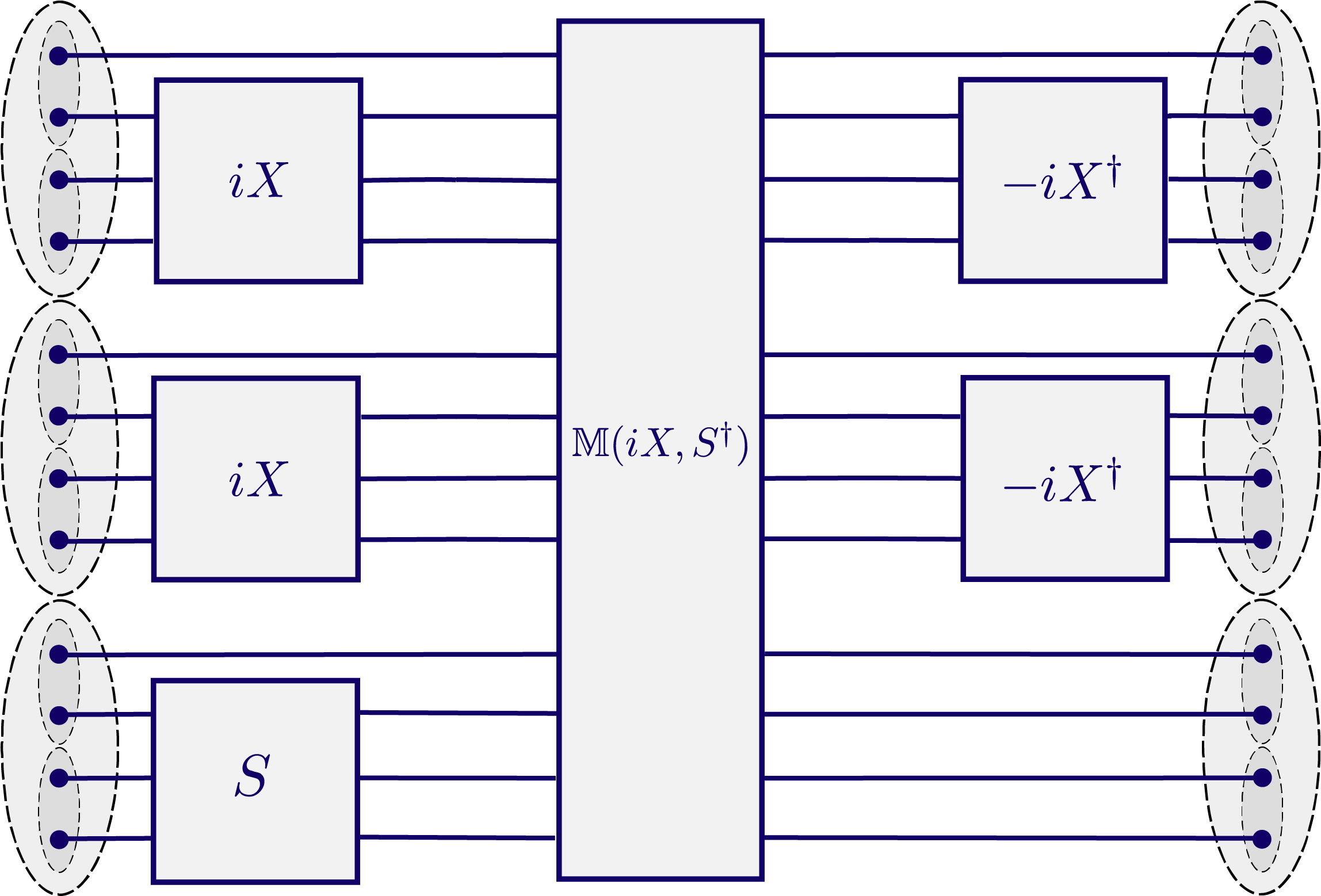}
    \caption{To compile the CCS gate with fewer braids, it is sufficient to add NOT gates, replaced by $iX$ and its Hermitian conjugate, and $S$ gate to the previously defined $\mathbb{M}(iX, S^\dagger)$ three-qubit gate.
    \label{fig:ccs}}
\end{figure}

A direct application of the topological gates previously introduced is the construction of the controlled-controlled-S (CCS) gate for all unitary $S$ up to the global phase determined by Eq. \eqref{eq:u(b)-formula}. The operationality of the target gate $S$ of this controlled-controlled gate is contingent upon both controlling qubits being in the state $\ket{1}$.
The implementation of the CCS gate implies the execution of the Deutsch gates $\mathbb{D}(\theta)$, which are intrinsically universal \cite{Deutsch1989}. The Deutsch gate is characterized by its target gate $S$, defined as:
\begin{align}
    S(\theta) &=
    \begin{pmatrix}
        \cos{\theta}   & -i\sin{\theta} \\
        -i\sin{\theta} & \cos{\theta}
    \end{pmatrix}
\end{align}
for any angle $\theta$.
A notable gate within this class is the Toffoli gate, renowned for its ability to compute any arbitrary Boolean function, thereby qualifying it as a universal reversible logic gate \cite{Toffoli1980}.\\
One way to compile the CCS gate is to combine $\mathbb{M}(I, S^\dagger)$ and  $\mathbb{M}(iX, S)$ since:
\begin{align}
    \text{CCS} &= \mathbb{M}(I, S^\dagger)\mathbb{M}(iX, S)\\
               &= \mathbb{M}(iX, S)\mathbb{M}(I, S^\dagger).
\end{align}
However, it is possible to construct a CCS gate with a single implementation of the $\mathbb{M}(R,S)$ gate. Initially, it should be noted that if we apply the $S$ gate on the target qubit prior to applying the $\mathbb{M}(iX,S^\dagger)$ gate, we should obtain a controlled-controlled gate which operates exclusively when both of the controlling qubits are in the $\ket{0}$ state. Consequently, to produce the desired CCS gate, NOT gates should be applied on the controlling qubits both before and after the $\mathbb{M}(iX,S)$ is applied. Given that such NOT gates occur symmetrically, we can set them to be $\pm iX$ then $\mp iX$ espectively, to eliminate the additional phase factor on the controlling qubits. The illustration of the CCS braid circuit is depicted in Fig. \ref{fig:ccs}. The corresponding truth table is detailed in Tab. \ref{tab:truth-table-CCS}. Therefore, the resulting matrix representation should be as follows:
\begin{align}
    \left(
    \begin{array}{cc|cc|cc|cc}
        1 & 0 & 0 & 0 & 0 & 0 & 0 & 0\\
        0 & 1 & 0 & 0 & 0 & 0 & 0 & 0\\
        \hline
        0 & 0 & 1 & 0 & 0 & 0 & 0 & 0\\
        0 & 0 & 0 & 1 & 0 & 0 & 0 & 0\\
        \hline
        0 & 0 & 0 & 0 & 1 & 0 & 0 & 0\\
        0 & 0 & 0 & 0 & 0 & 1 & 0 & 0\\
        \hline
        0 & 0 & 0 & 0 & 0 & 0 & \multicolumn{2}{c}{\multirow{2}{*}{\huge $S$}}\\
        0 & 0 & 0 & 0 & 0 & 0 & & 
    \end{array}
    \right)
\end{align}

\begingroup
    \begin{table}[h]
    \centering
    \caption{\label{tab:truth-table-CCS} Truth table of $CCS$ Gate.}
    \begin{ruledtabular}
    \begin{tabular}{cccccc}
        \multicolumn{3}{c}{Input} & \multicolumn{3}{c}{Output}\\
        \hline
        qubit 1   & qubit 2   & qubit 3      & qubit 1   & qubit 2   & qubit 3 \\
        \hline
        $\ket{0}$ & $\ket{0}$ & $\ket{\eta}$ & $\ket{0}$ & $\ket{0}$ & $\ket{\eta}$ \\
        $\ket{0}$ & $\ket{1}$ & $\ket{\eta}$ & $\ket{0}$ & $\ket{1}$ & $\ket{\eta}$ \\
        $\ket{1}$ & $\ket{0}$ & $\ket{\eta}$ & $\ket{1}$ & $\ket{0}$ & $\ket{\eta}$ \\
        $\ket{1}$ & $\ket{1}$ & $\ket{\eta}$ & $\ket{1}$ & $\ket{1}$ & $S\ket{\eta}$
    \end{tabular}
    \end{ruledtabular}
    \end{table}
\endgroup

\subsubsection{Numerical simulation of the controlled-controlled $\pm iX$ gate}
\label{sec:controlled-injection-ccx}
To obtain a special unitary version of the Toffoli gate, let us consider $S=\pm iX$. Initially, we should identify the compilation of $\pm \id$ and $\pm iX$ such that the weaving process commences from the upper strand and concludes at the lower strand. We should also find the target $S=\pm iX$ gate acting on three strands, initiating the weaving process from the upper strand and terminating at the same strand. By employing a brute-force approach, we derived the weaving sequences, as explicitly presented in Table \ref{tab:iToffoli}. It is important to note that these sequences are not the only possible optimal sequences. However, their combination yields the most accurate $\pm i$Toffoli gate.\\

\begingroup
\squeezetable
    \begin{table*}
    \centering
    \caption{\label{tab:iToffoli} Braid approximation of the necessary three-strand gates to build the $\pm i$Toffoli gate with overall error of $\toffolioverallerror$ and leakage $\epsilon_L = \toffolileakageerror$}.
    \begin{ruledtabular}
    \begin{tabular}{lllcc}
        CCS Gates & $\pm i$Toffoli Gates & Weave sequence & Length & Error\\
        \hline
        &&&\\
        R & $iX$ & 
        \scriptsize{$\sigma_1^{1}\sigma_1^{2}\sigma_2^{2}\sigma_1^{-2}\sigma_2^{2}\sigma_1^{-2}\sigma_2^{-2}\sigma_1^{-2}\sigma_2^{4}\sigma_1^{-2}\sigma_2^{-2}\sigma_1^{-2}\sigma_2^{2}\sigma_1^{-2}\sigma_2^{2}\sigma_1^{-2}\sigma_2^{2}\sigma_1^{2}\sigma_2^{2}\sigma_1^{-2}\sigma_2^{-2}\sigma_1^{2}\sigma_2^{-2}\sigma_2^{1}$} 
        & 48 & $8.55\times 10^{-4}$\\
        I & $\id$ 
        & \scriptsize{$\sigma_1^{1}\sigma_1^{2}\sigma_2^{2}\sigma_1^{2}\sigma_2^{2}\sigma_1^{-2}\sigma_2^{2}\sigma_1^{4}\sigma_2^{2}\sigma_1^{-2}\sigma_2^{2}\sigma_1^{2}\sigma_2^{-2}\sigma_1^{4}\sigma_2^{4}\sigma_1^{2}\sigma_2^{-2}\sigma_1^{2}\sigma_2^{4}\sigma_1^{2}\sigma_2^{-1}$} & 48 & $1.51\times 10^{-3}$\\
        S & $iX$ 
        & \scriptsize{$\sigma_2^{1}\sigma_2^{4}\sigma_1^{-2}\sigma_2^{-2}\sigma_1^{-2}\sigma_2^{-4}\sigma_1^{-2}\sigma_2^{2}\sigma_1^{2}\sigma_2^{-4}\sigma_1^{2}\sigma_2^{4}\sigma_1^{-2}\sigma_2^{4}\sigma_1^{-2}\sigma_2^{-4}\sigma_1^{-2}\sigma_2^{-2}\sigma_2^{-1}
        $} 
        & 48 & $8.55\times 10^{-4}$\\
        NOT & $iX$ 
        & \scriptsize{$\sigma_2^{1}\sigma_2^{4}\sigma_1^{-2}\sigma_2^{-2}\sigma_1^{-2}\sigma_2^{-4}\sigma_1^{-2}\sigma_2^{2}\sigma_1^{2}\sigma_2^{-4}\sigma_1^{2}\sigma_2^{4}\sigma_1^{-2}\sigma_2^{4}\sigma_1^{-2}\sigma_2^{-4}\sigma_1^{-2}\sigma_2^{-2}\sigma_2^{-1}
        $} 
        & 48 & $8.55\times 10^{-4}$\\
        &&
    \end{tabular}
    \end{ruledtabular}
    \end{table*}
\endgroup

\begin{figure}[t]
     \centering
     \begin{subfigure}[b]{\subfigleng}
         \centering
         \includegraphics[width=\textwidth]{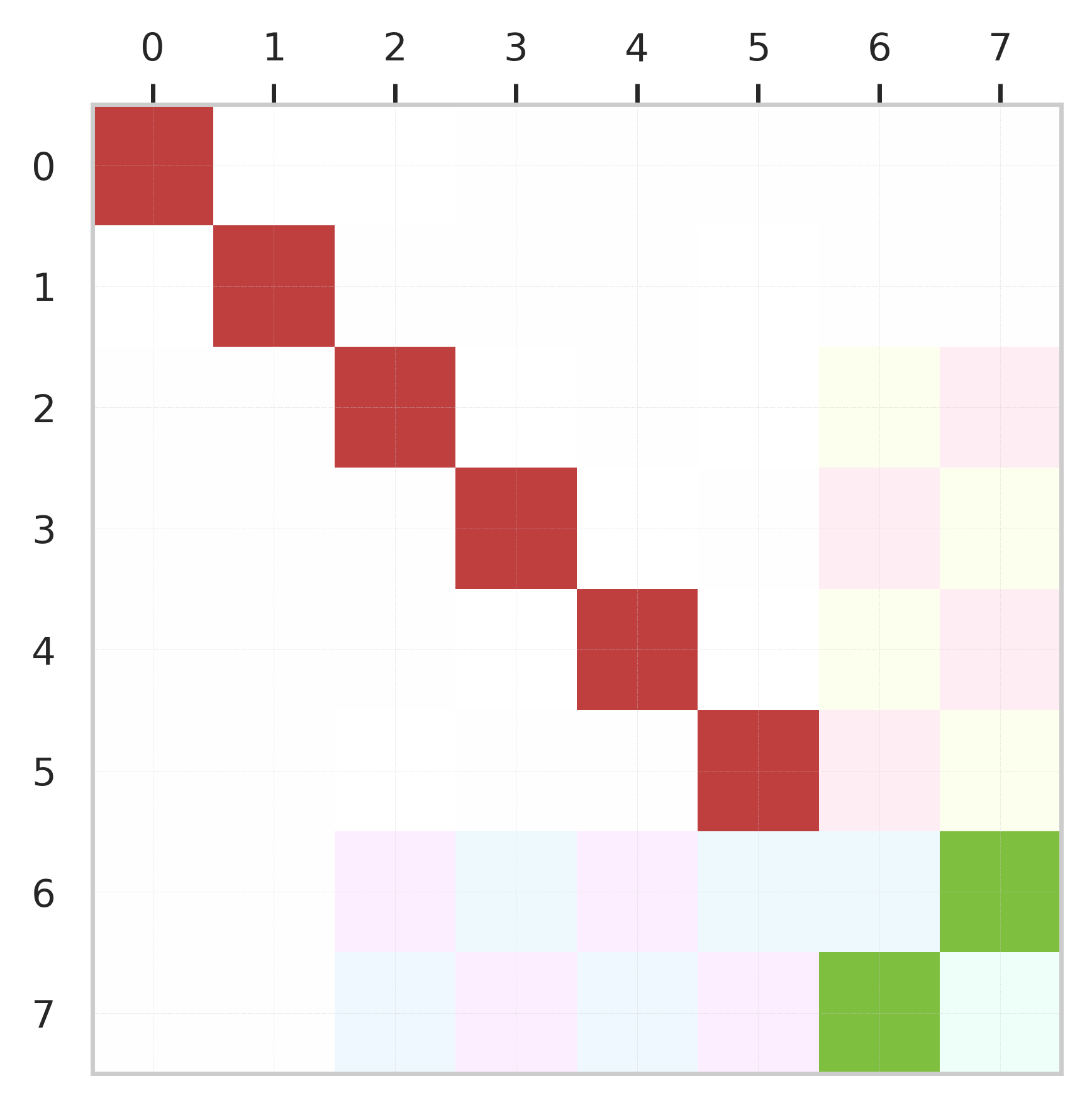}
         \caption{}
         \label{fig:iToffoli-sim}
     \end{subfigure}
     \begin{subfigure}[b]{\subfigleng}
         \centering
         \includegraphics[width=\textwidth]{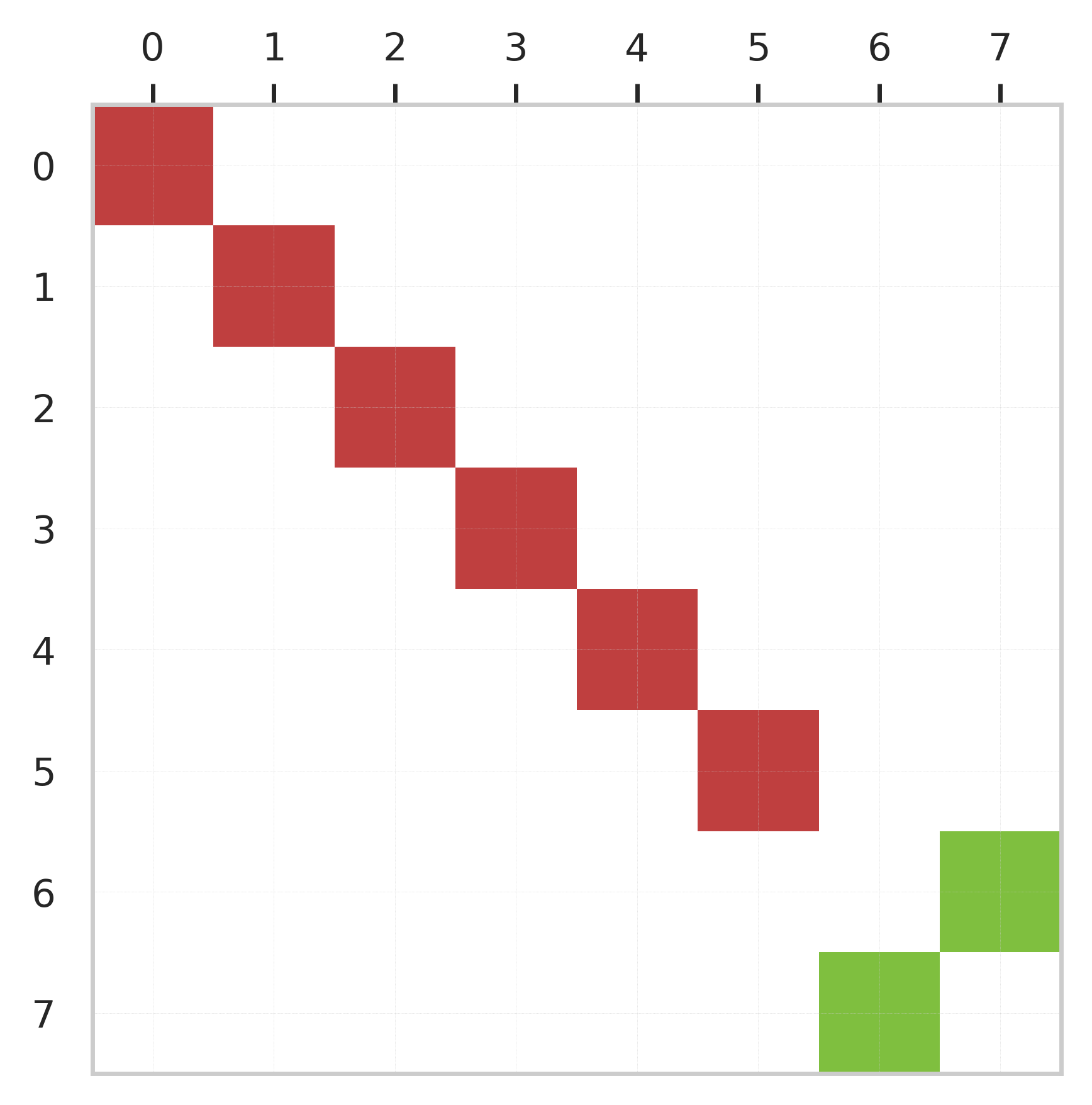}
         \caption{}
         \label{fig:iToffoli-exact}
     \end{subfigure}
        \caption{(a) Matrix representation of the approximated $i$Toffoli matrix in the basis of the computational states. It represents $i$Toffoli up to $1.02\times 10^{-4}$ distance error. (c) Exact $i$Toffoli matrix representation. The color map is depicted in Fig. \ref{fig:scale}.}
        \label{fig:iToffoli-simulation}
\end{figure}

Subsequently, the three-strand gates should be translated into a six-strand braiding circuit, as depicted in Fig. \ref{fig:three-qubit-gate}. The relevant braid matrices are then calculated, following the method employed in previous sections. Ultimately, we obtain the representation of the approximated $\mathbb{M}(I,iX)$ gate, as illustrated in Fig. \ref{fig:iToffoli-simulation}. The overall distance of the approximated $i$Toffoli gate is approximately $\toffolioverallerror$. However, the error solely on the controlled gate is about $\toffolitargeterror$. Lastly, the amount of leakage reaches $\toffolileakageerror$.\\

\section{Discussion}
\label{sec:discussion}
A distinctive feature of the CCS gate, compiled using the controlled-injection method, is its requirement for fewer than five two-qubit gates. Specifically, only four two-qubit gates are required: the initialization gates $R$ and $R^\dagger$, and the injection gates $I$ and $I^\dagger$.  At first glance, this seems to contradict the theorem etablished in \cite{PhysRevA.88.010304} which asserts that five two-qubit gates are necessary to implement the Toffoli gate.
Nevertheless, it is crucial to acknowledge that this theorem relies on the Hilbert space of qubits, whereas the fusion space of anyons exhibits a distinct structure with additional dimensions. The non-computational sub-space assumes a pivotal role in the functioning of initialization gates, facilitating the formation of $\mathbb{M}(R, S)$ gates. In summary, the inclusion of non-computational states makes it feasible to diminish the number of necessary two-qubit gates for the construction of a controlled three-qubit gate.
Furthermore, it is instructive to examine the disparities between the controlled-injection method and the standard decomposition method of CCS gates, detailed in Appendix \ref{app:decomposition-method}. Primarily, it is worth noting that the logic tables of $\mathbb{M}(R,S)$ gates can additionally contribute to the diversity of quantum logic gates. Additionally, the performance of these two methods are compared in terms of four aspects: length (number of elementary braids), accuracy, leakage, and ease of implementation. A numerical comparative study of Toffoli gate compilation using both methods is depicted in Tab. \ref{tab:comparison}. In our study, we chose the braid length to be 48 to give a reasonably acceptable approximation of the target gates up to an order of magnitude of $10^{-3}$, without requiring unreasonable computational resources and long braid sequences.
\begingroup
    \begin{table}[t]
    \centering
    \caption{
    \label{tab:comparison}
    Comparative table of the controlled-injection method and the decomposition method applied on the Toffoli CCX gate as shown in Section \ref{sec:controlled-injection-ccx} and Appendix \ref{app:decomposition-method} respectively. The parameters considered in this comparison include the number of required three-anyon gate approximations, the total number of braids needed to compile the gate (referred to as the required braid length), and the number of sequential braids in each gate (referred to as the depth). Additionally, numerical accuracy and leakage were also evaluated. In this context, $L$ denotes the braid length of single three-anyon gates. For the purpose of this numerical study, $L$ was set to 48. The error statistics in the target gate and leakage values were computed over all possible combinations of the optimal weaves.}
    \begin{ruledtabular}
    \begin{tabular}{lcc}
                                 & Decomposition              & Controlled-Injection   \\
        \hline
        Two-qubit gates          & 7                          & 4                      \\
        Three-anyon gates        & 3                          & 3                      \\
        Length                   & 30L+32                     & 25L                    \\
        Depth                    & 30L+32                     & 22L                    \\
        Best error               & $\toffolidecooverallerror$ & $\toffolioverallerror$ \\
        Leakage of the best      & $\toffolidecoleakageerror$ & $\toffolileakageerror$ \\
        \hline
        Avg error in target      & $2.32\times 10^{-3}$       & $2.35\times 10^{-3}$   \\
        Min error in target      & $1.76\times 10^{-3}$       & $8.54\times 10^{-4}$   \\
        Max error in target      & $3.16\times 10^{-3}$       & $3.34\times 10^{-3}$   \\
        \hline
        Avg leakage              & $1.75\times 10^{-6}$       & $2.031\times 10^{-6}$  \\
        Min leakage              & $3.93\times 10^{-7}$       & $3.59\times 10^{-7}$   \\
        Max leakage              & $4.09\times 10^{-6}$       & $3.99\times 10^{-6}$
    \end{tabular}
    \end{ruledtabular}
    \end{table}
\endgroup
Indeed, the controlled-injection method exhibits a shorter compilation length compared to the standard decomposition method. Assuming three-anyon gates of identical lengths, i.e., compiled with the same number $L$ of braids and consuming the same amount of time to search for the best compilation, the standard decomposition method would require $6L$ braids for each controlled two-qubit gate and $16$ for each SWAP gate, resulting in a total of $(30L + 32)$ braids.
In contrast, the controlled-injection method requires $4L$ braids for each three-anyon gate, i.e., $20L$ braids for the $\mathbb{M}(iX,S)$ gate, and an additional $5L$ braids to compile the extra single-qubit gates, yielding a total of $25L$ braids.
When comparing them according to depth, which is the number of sequential braids, we get $(30L+32)$ for decomposition and $22L$ for the controlled-injection method. Generally, depth is more significant in terms of compilation time. Therefore, the controlled-injection method provides gates that are at least $27\%$ times shorter.
Secondly, the controlled-injection method is as accurate as the standard decomposition method, or at least within the same order of magnitude, given that the composing gates are compiled with similar accuracies. Asymptotically, both methods should exhibit behavior identical to that of the $i$Toffoli gate. This demonstrates the efficacy of the controlled-injection method in terms of accuracy, making it a viable alternative to the standard decomposition method. 
Thirdly, both methods were found to yield a numerically similar amount of leakage. Generally, the injection and initialization gates are the primary sources of leakage. Given that we employ injection braids of the same accuracy and leakage in both methods, a significant difference in the amount of leakage between the two methods is not anticipated. This observation underscores the comparable performance of the controlled-injection method and the standard decomposition method in terms of leakage.
Lastly, the implementation of the controlled-injection method may present certain challenges, as it involves braids of multiple anyons (four anyons simultaneously), as opposed to weaving a single particle through the entire circuit. The manipulation of numerous anyons could pose significant technological hurdles \cite{Simon2006}. Conversely, the standard decomposition can be partially transformed at the level of CNOT gates into a weaving sequence of a single particle. This transformation is feasible due to the $FPF^{-1}$ braid circuit introduced in \cite{Hormozi2007}. Thus, while the controlled-injection method offers certain advantages, its practical implementation may require overcoming additional complexities.

\section{Conclusion}

In this study, we present an efficient topological quantum circuit model for compiling controlled-controlled gates within the Fibonacci anyon model, a versatile framework for universal quantum computing. We elevate the conventional numerical brute-force method by incorporating novel algebraic relations, such as similar braid sequences, and leveraging the symmetry by Hermitian target gates. Complemented by numerical methods like distributed computing, we achieve optimal approximations of single-qubit gates with three Fibonacci anyons.

In the concluding phase, we present a class of conditional three-qubit gates designed for approximating controlled logic operations, including fundamental gates like Deutsch and Toffoli. These gates are seamlessly compatible with the inherent structure of the Fibonacci model. A comparative analysis between the conventional decomposition method of three-qubit gates and the novel controlled-injection approach reveals that the latter allows for a more concise compilation, achieving shorter lengths and depths in implementing controlled-controlled gates, while preserving a comparable degree of accuracy and leakage.

Remarkably, the introduced controlled three-qubit gates are decomposed into four, rather than the conventional five, two-qubit gates, as stipulated by the decomposition theorem. This reduction is made possible by harnessing the non-computational topological states within the fusion space. The controlled-injection method leverages on the unique logical attributes of the Fibonacci anyon model, presenting a potential template for analogous strategies across diverse anyon models or with an increased number of qubits.

\appendix
\section{Similar braid sequences}
\label{app:FBF}
The similarity relation between the braid sequences of three anyons starting with $\sigma_1$ and $\sigma_2$ has a practical use in optimizing the search for the best approximation of a given target unitary gate. The main observation that leads to similar braid sequences is the fact that applying $\sigma_1$ and $\sigma_2$ braid operations on the state $\ket{((a,b)_i,c)_j}$ of three anyons $a$, $b$ and $c$ is equivalent to applying $\sigma_2$ and $\sigma_1$, respectively, on the state $\ket{(c, (b, a)_i)_j}$, as another observer may choose to look at the same set of anyons from the opposite side. In other words, $\sigma_1$ transforms to $\sigma_2$ when rotating the surface $180^\circ$. Namely,

\begin{widetext}
\begin{align}
    \left|\bra{((a, b)_{i'}, c)_j}\sigma_1\ket{((a,b)_i, c)_j}\right|^2 &= \left|\bra{(c, (b, a)_{i'})_j}\sigma_2\ket{(c,(b, a)_i)_j}\right|^2\\
    \left|\bra{((a, b)_{i'}, c)_j}\sigma_2\ket{((a,b)_i, c)_j}\right|^2 &= \left|\bra{(c, (b, a)_{i'})_j}\sigma_1\ket{(c,(b, a)_i)_j}\right|^2
\end{align}
\end{widetext}

This is true because, topologically, the three following operations are equivalent:
\begin{align*}
    \sigma_1 \sigma_2 \sigma_1 \ket{((a,b)_i, c)_j} 
    &\equiv \sigma_2 \sigma_1 \sigma_2 \ket{((a,b)_i, c)_j},\nonumber\\
    &\equiv R_{ab} R_{ic} \ket{((a,b)_i, c)_j},
\end{align*}
and
\begin{align}
    R_{ab} R_{ic} \ket{((a,b)_i, c)_j} &= R_{ab}^i R_{ic}^j \ket{(c,(b,a)_i)_j}.
\end{align}
These three operations do nothing but rotate the frame of anyons $180^\circ$.
The first equivalence relation of the three is is one of the Artin relations of the braid group.\\
Now, let's define the operator $\Gamma$ such as:
\begin{align*}
    \Gamma &= \sigma_1 \sigma_2 \sigma_1\\ 
           &= \sigma_2 \sigma_1 \sigma_2.
\end{align*}
It is easy to see that
\begin{align}
    \sigma_1 &= \Gamma^\dagger \sigma_2 \Gamma \label{eq:gamma1}, \\
    \sigma_2 &= \Gamma^\dagger \sigma_1 \Gamma \label{eq:gamma2}.
\end{align}
In general, a braid sequence $\text{Braid}(\sigma_1, \sigma_2)$ applied on three anyons takes the form:
\begin{align}
    \text{Braid}(\sigma_1, \sigma_2) = \sigma_1^{p_n}\sigma_2^{p_{n-1}}\cdots \sigma_1^{p_{2}}\sigma_2^{p_{1}}
\end{align}
where $p_i$ can be any integer and $n\geq 2$. Therefore, the permutation between $\sigma_1$ and $\sigma_2$ yields the braid sequence $\overline{\text{Braid}}(\sigma_1, \sigma_2)$ such that
\begin{align}
    \overline{\text{Braid}}(\sigma_1, \sigma_2) &= \text{Braid}(\sigma_2, \sigma_1) = \sigma_2^{p_n}\sigma_1^{p_{n-1}}\cdots \sigma_2^{p_{2}}\sigma_1^{p_{1}}
\end{align}
The previous relations Eq. \eqref{eq:gamma1} and Eq. \eqref{eq:gamma2} imply that $\text{Braid}$ and $\overline{\text{Braid}}$ are related by the similarity relation:
\begin{align}
    \overline{\text{Braid}}(\sigma_1, \sigma_2) = \Gamma^\dagger \text{Braid}(\sigma_1, \sigma_2) \Gamma.
\end{align}
This similarity relation can be deduced immediately by substitution.

\section{Distance symmetry induced by Hermitian target}
\label{app:proposition}
\begin{proposition}
Let $(G,\mathcal{D})$ be a metric space such that the set $G$ forms a group and $\mathcal{D}$ is a 
\textit{bi-invariant} metric. Then, for all $a,b \in G$, if $b$ is its own inverse (i.e., $b=b^{-1}$), it follows that $\mathcal{D}(a,b)=\mathcal{D}(a^{-1},b)$.
\end{proposition}
A metric $\mathcal{D}$ is defined to be \textit{bi-invariant} if, for all $a, b, c \in G$, we have $\mathcal{D}(a, b) = \mathcal{D}(ac, bc) = \mathcal{D}(ca, cb)$ \cite{Huynh2009}. Given that the set of special unitary matrices forms a group and the spectral distance is bi-invariant, let $H$ be a target unitary that is a Hermitian matrix. If $\mathcal{D}(B, H)$ is known for a given braid matrix $B$, then $\mathcal{D}(B^\dagger, H)$ holds the same value.

\section{Approximated Toffoli gate with decomposition method}
\label{app:decomposition-method}
The CCS gate can be decomposed to at least five two-qubit controlled operations \cite{Barenco1995, PhysRevA.88.010304} as shown in Fig. \ref{fig:ccu}. Namely, for any unitary $\text{U} \in \text{U(2)}$,
\begin{align}
    \controlled{\controlled{\text{U}}} =
    &(\swapp \otimes \id)
    (\id \otimes \controlled{\sqrt{\text{U}}})
    (\swapp \otimes \id)\nonumber                       \\
    &(\controlled{\nott} \otimes \id)
    (\id \otimes \controlled{\sqrt{\text{U}}^\dagger})
    (\controlled{\nott} \otimes \id)\nonumber
    (\id \otimes \controlled{\sqrt{\text{U}}})
\end{align}
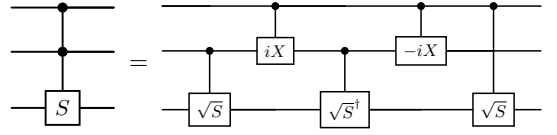
\begin{figure}[t]
    \begin{adjustbox}{width=0.18\linewidth}
    \begin{quantikz}
        & \ctrl{1} & \qw \\
        & \ctrl{1} & \qw \\
        & \gate{S} & \qw
    \end{quantikz}
    \end{adjustbox}
    =
    \begin{adjustbox}{width=0.6\linewidth}
    \begin{quantikz}
        & \qw             & \ctrl{1}  & \qw                     & \ctrl{1}   & \ctrl{2}        & \\
        & \ctrl{1}        & \gate{iX} & \ctrl{1}                & \gate{-iX} & \qw             & \\
        & \gate{\sqrt{S}} & \qw       & \gate{\sqrt{S}^\dagger} & \qw        & \gate{\sqrt{S}} & 
    \end{quantikz}
    \end{adjustbox}
    
    \caption{Controlled-Controlled-S (CCS) gate can be systematically decomposed into a sequence of five two-qubit controlled gates. This sequence comprises two Controlled-NOT (CNOT) gates and two controlled gates that implement the square root of the S gate.
             \label{fig:ccu}}
\end{figure}
\begin{figure}[t]
    \centering
    \includegraphics[width=\linewidth]{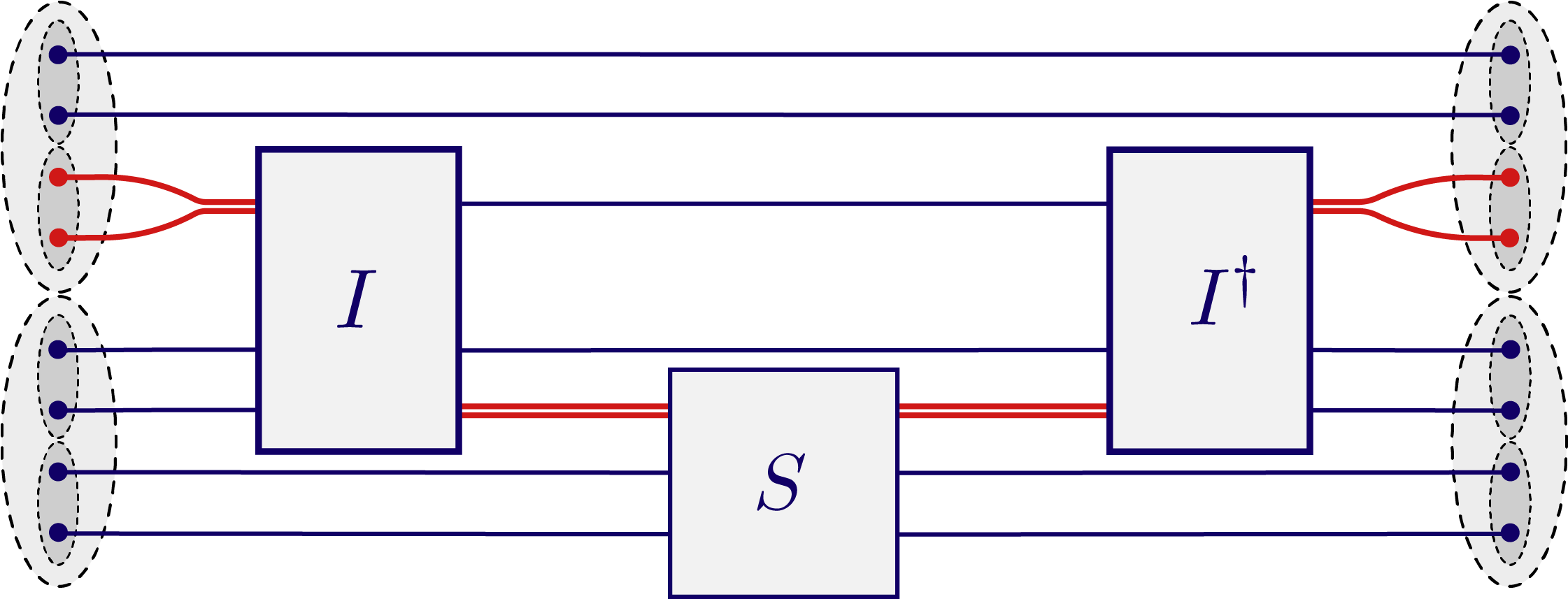}
    \caption{In the injection method, three-strand gates can implement two-qubit gates. Notice that each lower pair of anyons in the same qubit should yield the vacuum if the state is $\ket{0}$ and yield an anyon if it is $\ket{1}$. Therefore, weaving a pair of anyons around the anyons of the controlled qubit and then returning to the initial position does not affect the state unless the state is $\ket{1}$. That is the spirit of a controlled gate.
    \label{fig:cu-gate}}
\end{figure}
The SWAP gate is needed because the arrangement of anyons does not allow two-qubit operation between non-neighboring three-anyons qubits without making non-trivial exchanges with the intermediary anyons. Therefore, to compile the CCS gate by decomposition, it is necessary to construct C$i$NOT, C$\sqrt{S}$ and SWAP gates with sufficient accuracy.

While the SWAP gate is trivial in the context of encoding qubits with groups of four anyons whose overall topological charge is $\vac$, the involved two-qubit controlled gates C$i$NOT and C$\sqrt{S}$ can be approximated by the injection method introduced in \cite{Bonesteel2005} as shown in Fig. \ref{fig:cu-gate}.

\subsubsection{Numerical simulation}
\begin{figure}[h]
     \centering
     \begin{subfigure}[b]{\subfigleng}
         \centering
         \includegraphics[width=\textwidth]{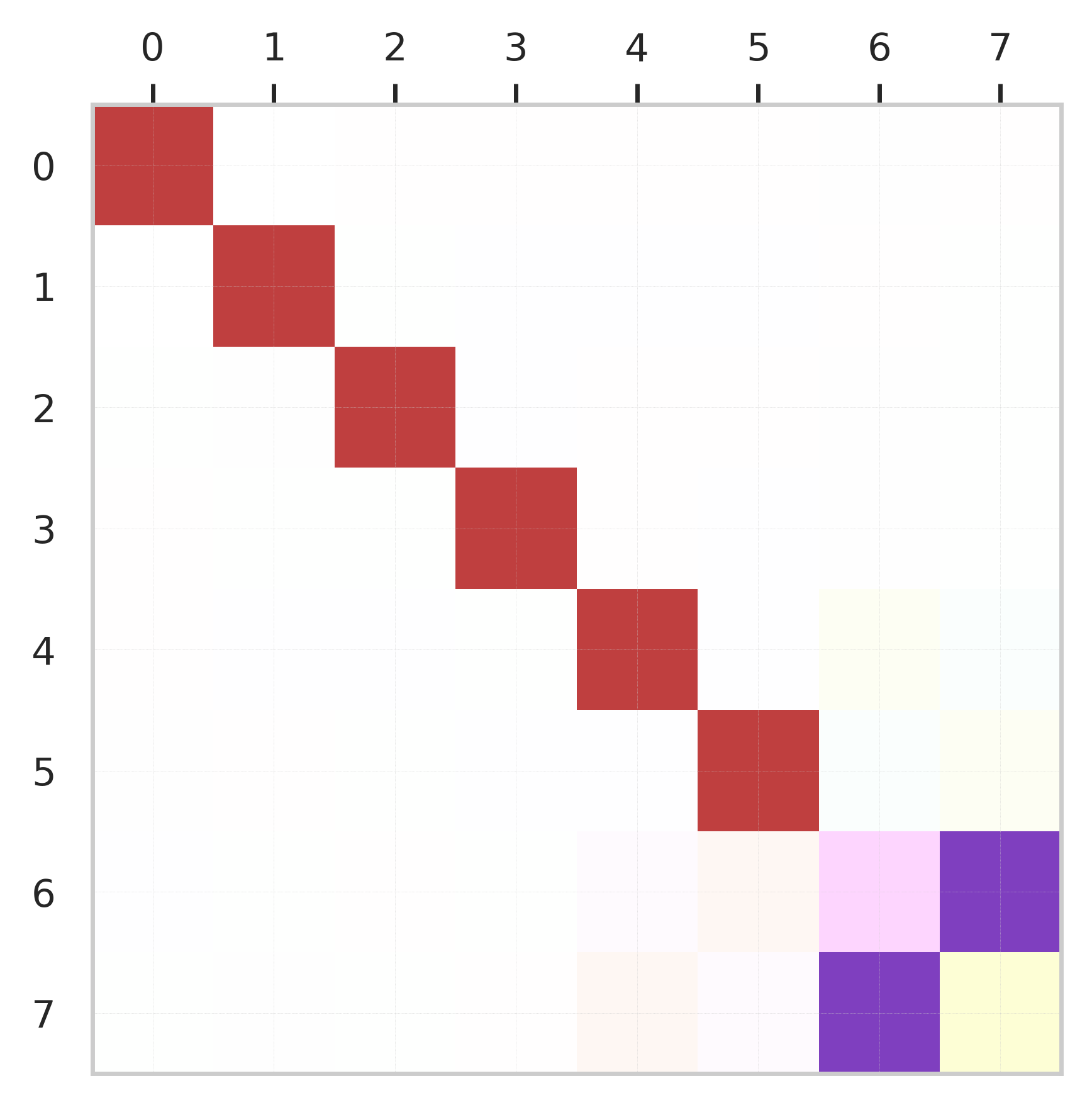}
         \caption{\label{fig:iToffoli-decompo-sim}}
     \end{subfigure}
     \begin{subfigure}[b]{\subfigleng}
         \centering
         \includegraphics[width=\textwidth]{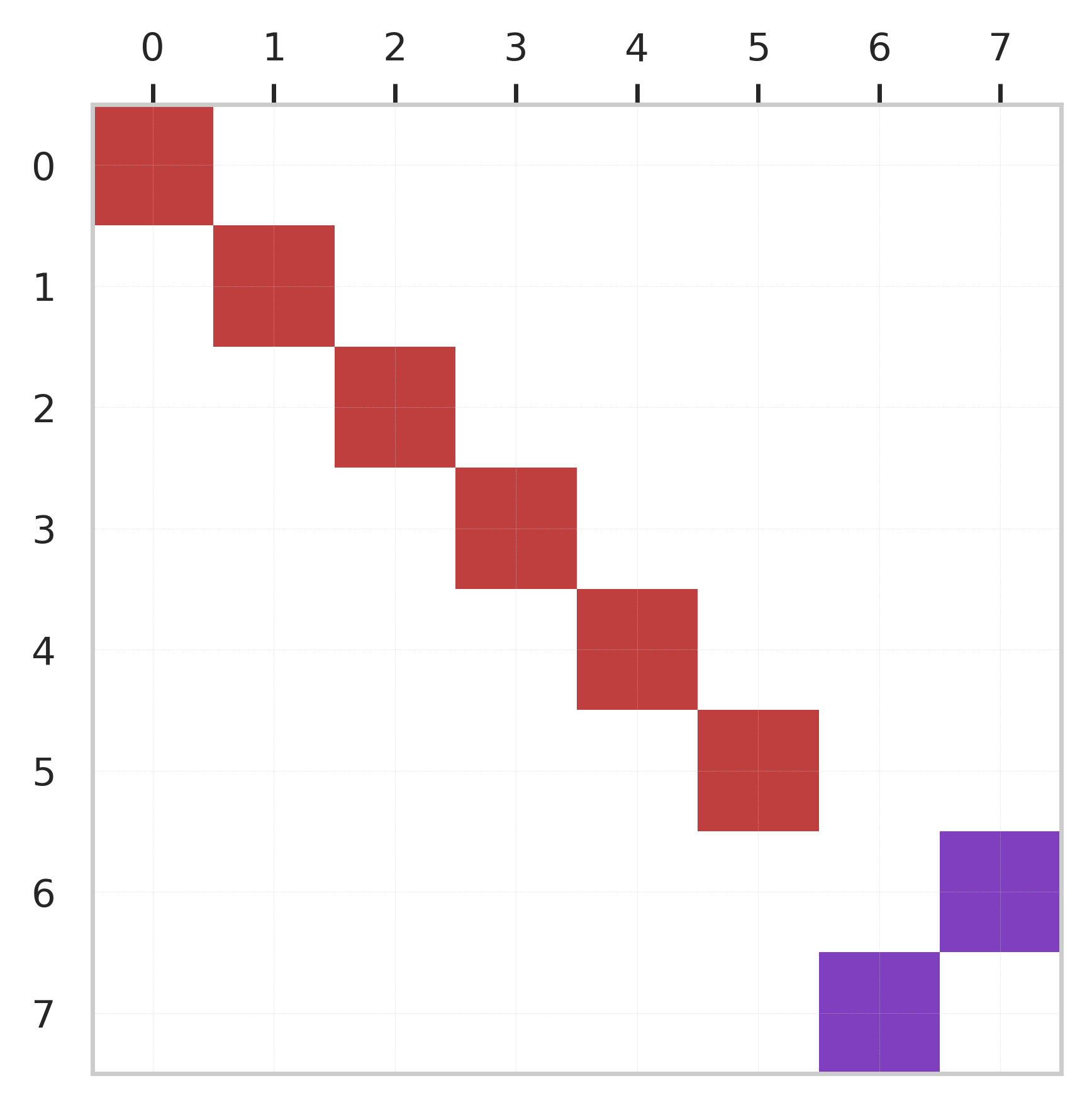}
         \caption{\label{fig:iToffoli-decompo-exact}}
     \end{subfigure}
        \caption{(a) Matrix representation of the approximated $-i$Toffoli gate, expressed in the computational basis. (b) Exact matrix representation of the $-i$Toffoli gate. The color map is depicted in Fig. \ref{fig:scale}.
        \label{fig:iToffoli-decompo-simulation}}
\end{figure}

\begingroup
\squeezetable
    \begin{table*}[t]
    \centering
    \caption{Braid approximation of the necessary three-strand gates to build the $i$Toffoli gate with the decomposition method. We use two different best braid sequences to approximate $I$ for CNOT and C$\sqrt{\text{NOT}}$ and optimize accuracy and leakage.
    \label{tab:iToffoli-decomposition}}
    \begin{ruledtabular}
    \begin{tabular}{lllcc}
        Gates & Target Gates & Weave sequence & Length & Error\\
        \hline
        &&&\\
        CNOT Injection & $\id$ & \scriptsize{$\sigma_1^{-1}\sigma_1^{-2}\sigma_2^{2}\sigma_1^{4}\sigma_2^{2}\sigma_1^{4}\sigma_2^{-2}\sigma_1^{2}\sigma_2^{-2}\sigma_1^{2}\sigma_2^{4}\sigma_1^{2}\sigma_2^{-2}\sigma_1^{2}\sigma_2^{-4}\sigma_1^{-2}\sigma_2^{-4}\sigma_1^{-2}\sigma_2^{-2}\sigma_2^{-1}$} & 48 & $1.51\times 10^{-3}$\\
        C$\sqrt{\nott}$ Injection & $\id$ & \scriptsize{$\sigma_1^{1}\sigma_1^{2}\sigma_2^{2}\sigma_1^{4}\sigma_2^{4}\sigma_1^{4}\sigma_2^{2}\sigma_1^{4}\sigma_2^{2}\sigma_1^{2}\sigma_2^{-2}\sigma_1^{2}\sigma_2^{-2}\sigma_1^{2}\sigma_2^{2}\sigma_1^{2}\sigma_2^{-2}\sigma_1^{2}\sigma_2^{-2}\sigma_1^{2}\sigma_2^{-1}$} & 48 & $1.51\times 10^{-3}$\\
        $\sqrt{\nott}$ & $-i\sqrt{iX}$ & \scriptsize{$\sigma_2^{-1}\sigma_1^{2}\sigma_2^{4}\sigma_1^{-2}\sigma_2^{-4}\sigma_1^{2}\sigma_2^{-2}\sigma_1^{2}\sigma_2^{4}\sigma_1^{-2}\sigma_2^{2}\sigma_1^{2}\sigma_2^{4}\sigma_1^{2}\sigma_2^{-2}\sigma_1^{4}\sigma_2^{4}\sigma_1^{2}\sigma_2^{-1}$} & 48 & $1.24\times 10^{-3}$\\
        NOT & $iX$ & \scriptsize{$\sigma_2^{1}\sigma_2^{4}\sigma_1^{-2}\sigma_2^{-2}\sigma_1^{-2}\sigma_2^{-4}\sigma_1^{-2}\sigma_2^{2}\sigma_1^{2}\sigma_2^{-4}\sigma_1^{2}\sigma_2^{4}\sigma_1^{-2}\sigma_2^{4}\sigma_1^{-2}\sigma_2^{-4}\sigma_1^{-2}\sigma_2^{-2}\sigma_2^{-1}$} & 48 & $8.55\times 10^{-4}$\\
        &&
    \end{tabular}
    \end{ruledtabular}
    \end{table*}
\endgroup
We should find first the compilation of $\pm \id$ such that it starts weaving from the upper strand and ends up in the lower strand. We need also to approximate $\pm iX$ and $\sqrt{\pm iX}$ gate acting on three strands starting weaving from the upper strand and ending up in the same strand. By brute-forcing, we find the required weaving sequences as shown in Tab. \ref{tab:iToffoli-decomposition}. The simulated $-i$Toffoli gate by decomposition is represented in Fig. \ref{fig:iToffoli-decompo-simulation}. The distance error of this approximated $-i$Toffoli gate is $\toffolidecooverallerror$. However, the error on the controlled gate only is about $\toffolidecotargeterror$. The leakage is computed as well and it is $\toffolidecoleakageerror$. These values exhibit a similar magnitude of approximation as those obtained using the controlled-injection method. However, the decomposition method generates longer braid sequences, as discussed in Section \ref{sec:discussion}.

\section*{Acknowledgments}
This document has been produced with the financial assistance of the European Union (Grant no. DCI-PANAF/2020/420-028), through the African Research Initiative for Scientific Excellence (ARISE), pilot programme. ARISE is implemented by the African Academy of Sciences with support from the European Commission and the African Union Commission.
We are grateful to the Algerian Ministry of Higher Education and Scientific Research and DGRST for the financial support.

\pagebreak
\bibliography{bibliography.bib}
\bibliographystyle{apsrev4-2}

\end{document}